\documentclass[twocolumn, 10pt, journal,letterpaper]{IEEEtran}
\usepackage{url}
\usepackage{graphicx} 
\usepackage{algorithm,setspace}
\usepackage[noend]{algpseudocode}
\usepackage{etex} 
\usepackage{tikz}
\usetikzlibrary{shapes,arrows}
\usepackage{pstricks}
\usepackage{pst-node,pst-blur}
\usetikzlibrary{arrows}
\usepackage{amsfonts}
\usepackage{tabularx}
\usepackage{subfigure}
\usepackage{amssymb}
\usepackage{booktabs}
\usepackage{multirow}
\usepackage{steinmetz} 
\usepackage{epsfig}
\usepackage{epstopdf}

\usepackage[noadjust]{cite}
\newtheorem{theorem}{Theorem}
\newtheorem{remark}{Remark}

\newtheorem{lemma}{Lemma}

\newtheorem{corollary}{Corollary}

\usepackage[cmex10]{amsmath} 
\usepackage[cmex10]{mathtools}

\newcommand\abs[1]{\left\lvert#1\right\rvert}
\newcommand\norm[1]{\left\lVert#1\right\rVert}

\allowdisplaybreaks

\begin{document} 
\title{Optimal Channel Estimation for Hybrid Energy Beamforming under Phase Shifter Impairments} 
\author{Deepak Mishra,~\IEEEmembership{Member,~IEEE,} and H\aa kan Johansson,~\IEEEmembership{Senior~Member,~IEEE}
	\thanks{D. Mishra and H. Johansson are with the Communication Systems Division of the Department of Electrical Engineering at the Link\"oping University, 581 83 Link\"oping, Sweden (emails: \{deepak.mishra, hakan.johansson\}@liu.se).}
	\thanks{This research work is funded by ELLIIT.}
	}

\maketitle

\begin{abstract} 
Smart multiantenna wireless power transmission can enable perpetual operation of energy harvesting (EH) nodes in the internet-of-things. Moreover, to overcome the increased hardware cost and space constraints associated with having large antenna arrays at the radio frequency (RF) energy source, the hybrid energy beamforming (EBF) architecture with single RF chain can be adopted. Using the recently proposed hybrid EBF architecture modeling the practical analog phase shifter impairments (API), we derive the optimal least-squares estimator for the energy source to EH user channel. Next, the average harvested power at the user is derived while considering the nonlinear RF EH model and a tight analytical approximation for it is also presented by exploring the practical limits on the API. Using these developments, the jointly global optimal transmit power and time allocation for channel estimation (CE) and EBF phases, that maximizes the  average energy stored at the EH user is derived in closed form. Numerical results validate the proposed analysis and present nontrivial design insights on the impact of API and CE errors on the achievable EBF performance. It is shown that the optimized hybrid EBF protocol with joint resource allocation yields an average performance improvement of $37\%$ over benchmark fixed allocation scheme.
\end{abstract} 

\IEEEpeerreviewmaketitle  
 
\section{Introduction}\label{sec:intro} 
Using large antenna array at the radio frequency (RF) source can enable perpetual operation of energy harvesting (EH) devices in internet of things (IoT)~\cite{ComMag} by compensating propagation losses through energy beamforming (EBF), or enhancing information capacity via multi-stream transmission~\cite{ArrayG}. Despite these potential merits, there are two practical fundamental bottlenecks: (1) larger physical size of the antenna arrays at usable RF frequencies~\cite{HBF-overview}, and (2) increased signal processing complexity because the digital precoding has to be applied over hundreds of antenna elements~\cite{HBF_survey}. Since, the multiantenna digital precoding is carried out at the baseband, each antenna element requires its own RF chain for the analog-to-digital  conversion, and subsequent baseband-to-RF up conversion, or vice-versa. This usage of one RF chain per antenna is very inefficient, both from  hardware monetary cost and energy consumption perspectives~\cite{ArrayG,HBF-overview,HBF_survey}. This has led to a growing research interest~\cite{HBF-overview,HBF_survey,HBF-CE-main,HBF-CE-conf1, HBF-Prac1, HBF-Prac2, HBF-CE-oth1,  ko2016channel,HBF-CE-oth5, HBF-CE-oth3,  HBF-CE-oth2} in the hybrid beamforming architectures, where all or part of the processing is based on analog beamforming which enables a substantially reduced number of RF chains in comparison to the  antenna count.

\subsection{State-of-the-Art}\label{sec:RW}  
We recall that accurate channel state information (CSI) is needed at the multiantenna energy source to maximize the array gains for meeting sustainable operation demand of EH IoT devices~\cite{Signal-WPT-TC17}. Different channel estimation (CE) schemes based on minimizing the least-squares (LS) error or linear minimum-mean square-error (MMSE) have been investigated in the literature for exploiting the fully digital energy beamforming  gains~\cite{CSI-SU-WET,ICASSP18,Signal-WPT-TC17}. \textcolor{black}{Keeping in mind the constraints of RF EH users, various limited feedback based CE protocols~\cite{New-One-Bit} and resource optimization  techniques~\cite{HIPT-TWC17} have also been recently studied. Further, the efficacy of received signal strength indicator (RSSI) feedback values based CE protocols has been lately investigated in~\cite{New-TSP,New-WiOpt}.} However, these fully-digital EBF works~\cite{Signal-WPT-TC17,CSI-SU-WET,ICASSP18,New-One-Bit,New-TSP,New-WiOpt,HIPT-TWC17} adopted an overly simplified linear rectification model for their investigation, which has been recently~\cite{Powercast-P1110,RFEH2008,ICC17Wksp,Sigmoidal-RFH,nonRFH2018,TWC17} shown to  perform poorly for the practical RF EH circuits. The detailed investigation on RF EH performance using the statistical CSI as conducted in~\cite{nonRFH2018,TWC17}  suggested that for an accurate characterization, a nonlinear EH model should be adopted during investigations.

In contrast to the multi-stream information transfer (IT) using multiantenna source, efficient RF energy transfer (RFET) involves the dynamic adjustment of the  beams from different antenna elements to focus most of the radiated RF power in the direction of an intended EH user. Also, it has been proved mathematically in~\cite{HBF-Prac1,HBF-Prac2} that by using two digitally controlled phase shifters (DCPS) for each antenna element the corresponding analog EBF with single RF chain can achieve exactly the same array gains as that of a fully digital system with each antenna element having its own RF chain.  Different from the digital beamforming works, a highly accurate CE process for implementing hybrid EBF is more challenging to realize because here the effective channel is the product of the random fading gain and analog beam selected~\cite{HBF-overview,HBF_survey,HBF-CE-main,HBF-CE-conf1, HBF-Prac1, HBF-Prac2, HBF-CE-oth1,  ko2016channel,HBF-CE-oth2, HBF-CE-oth3,  HBF-CE-oth5}. An adaptive compressed sensing (CS) based CE algorithm was proposed in~\cite{HBF-CE-oth1} for a hybrid analog-digital multiple-input-multiple-output (MIMO) system. Considering the multi-user hybrid beamforming system, a minimum mean-square  error (MMSE) approach was developed in~\cite{HBF-CE-main} to estimate the effective channel. Joint least-squares (LS) based CE and analog beam selection algorithm was proposed in~\cite{ko2016channel} for an uplink (UL) multiuser hybrid beamforming system. In contrast to these narrow band systems facing flat fading, CE algorithms for a single user multi-carrier hybrid MIMO system was investigated in~\cite{HBF-CE-conf1} using both  LS  and  CS approaches. More recently, a new  CE approach for hybrid architecture-based wideband millimeter wave systems was proposed in~\cite{HBF-CE-oth5} using the  sparse  nature  of frequency-selective channels. However, as obtaining full-dimensional instantaneous CSI  is  difficult  due  to  much lesser  RF  chains  than the antenna elements, a low-complexity hybrid  precoding  approach was investigated in~\cite{HBF-CE-oth3}  that involves  beam  searching in  the  downlink (DL) and  the analog  precoder codeword index feedback  in  the  UL. To alleviate the high hardware cost in complicated signaling procedure, a single-stage feedback scheme exploiting the second-order  channel  statistics for designing the digital  precoder and using the feedback  only for the analog beamforming was proposed in~\cite{HBF-CE-oth2}. Here, it is worth noting that these works~\cite{HBF-overview,HBF_survey,HBF-CE-main,HBF-CE-conf1, HBF-Prac1, HBF-Prac2, HBF-CE-oth1,  ko2016channel,HBF-CE-oth2, HBF-CE-oth3,  HBF-CE-oth5} focusing on multi-stream IT for efficient spatial multiplexing using limited RF chains  at the source or user, did not investigate the joint optimal CE protocol and resource allocation to maximize the EBF gains under hardware impairments.

\subsection{Motivation, Novelty, and Scope}\label{sec:motiv-novel-scope}
Analog EBF can address the hardware cost and space constraints in practically realizing efficient RFET from a large antenna array~\cite{HBF_survey}. However, due to the usage of low-cost hardware and low-quality RF components for the ubiquitous deployment of EH devices in IoT and for making large antenna array systems economically viable, the performance of these energy sustainable systems is more  prone to the RF imperfections caused by practical phase shifters (PS) and lossy combiners~\cite{schenk2008RF,bakr2009impact,PS1,PS2}. This may result in a significant EH performance degradation due to the underlying practical analog phase-shifter impairments (API). Recently, an API model was introduced in~\cite{SPL} for investigating the efficacy of MMSE-based CE for hybrid EBF over Rayleigh channels, while assuming a linear EH model. \textcolor{black}{However, this linear rectification model is only suitable when the received signal power levels are very low~\cite{ICC17Wksp,nonRFH2018}.} In contrast, here we aim at investigating the degradation in the hybrid EBF gains as compared to a fully digital architecture~\cite{CSI-SU-WET,ICASSP18} for the practical  multiple-input-single-output (MISO) RFET~\cite{Signal-WPT-TC17} over Rician fading channels~\cite[Ch 2.2]{simon2005digital}, while adopting a more refined EH model.  Rician fading is important as it incorporates the strong line-of-sight (LoS) components over RFET  links~\cite{TWC17}.  \textcolor{black}{Though the hybrid architecture can help in realizing significant monetary cost and energy consumption reduction due to the usage of a single RF chain, it is prone to hardware imperfections, such as phase offset errors between different DCPS pairs along with differences in their amplitude gains. However, these performance losses due to practical API, whose affect is characterized in this work, can be overcome by considering the proposed jointly optimal time and power allocation for the CE and RFET sub-phases.}  Further, noting  the energy constraints of an  EH user, we present a green transmission protocol involving optimal LS-based CE, which  does not require any prior knowledge on channel  statistics.

\textit{\textcolor{black}{To our best knowledge, the joint impact API, CE errors, and nonlinear rectification efficiency on the optimized average stored energy at single antenna EH user due to hybrid EBF during  MISO RFET  over Rician channels has not been investigated yet.}} Moreover, the existing works~\cite{Signal-WPT-TC17,CSI-SU-WET,ICASSP18,New-One-Bit,New-TSP,New-WiOpt,HIPT-TWC17} on resource allocation for optimizing the digital EBF  performance under CE errors,  considered an overly simplified linear RF EH model and presented either suboptimal or numerical solutions. In contrast, we focus on obtaining analytical insights on the joint design to optimally allocate resources between CE and RFET  phases. The major challenge is to obtain closed-form solution for the nonconvex stored energy maximization problem. 

The scope of this work involves the characterization of practical efficacy of hybrid  EBF  having a common single RF chain for a large array of antenna elements. The nontrivial outcomes and observations of this work for a single user DL wireless RFET  scenario can be extended to multiuser simultaneous wireless information and power transfer (SWIPT) applications~\cite{ComMag,TWC17}. Also, the adopted API model and optimal LS-based CE protocol for EBF with a single RF chain at multiantenna source can be extended to address the demands of hybrid EBF architectures with multiple RF chains. Further, the closed-form expressions for the joint design shed key insights on an efficient utilization of the available resources for maximizing the achievable array gains. \textcolor{black}{Lastly, with the latest developments of the low-power circuits capable of harvesting power from  millimeter wave energy signals~\cite{mmW_EH_imp2}, the proposed optimal CE and hybrid EBF designs can be used for sustainable high frequency indoor applications with much less bulkier power beacon.}

\subsection{Key Contributions and Notations}\label{sec:contrib}
The key contribution of this work is five fold. 
\begin{itemize} 
	\item We present a novel joint CE and energy optimization framework for maximizing the hybrid EBF efficiency during the RFET  over Rician channels while incorporating API at the multiantenna source and nonlinear rectification operation of RF EH unit at the user. The considered system model and the hybrid EBF architecture are presented in Section~\ref{sec:model}.	
	\item Global optimal LS estimator (LSE) for the effective channel, involving the product of channel vector and analog EBF design, is obtained in Section~\ref{sec:CE} while considering the impact of API. The key statistics for the LSE, involving analog and digital channel estimators, are derived along with their respective practically-motivated tight analytical approximations.
	\item Using this API-affected LSE, the average received RF power analysis is carried out in Section~\ref{sec:approx} while considering the nonlinear rectification operation in the practical RF EH circuits. Tight closed-form approximations for the average harvested power and the stored energy at the EH user after replenishing the consumption in CE are also derived in Section~\ref{sec:analysis} for the MISO RFET over the Rician fading channels, both with and without  CE errors.
	\item Green transmission protocol involving the joint optimal time and power allocation (PA) for CE and RFET sub-phases to maximize the stored energy at the EH user is investigated in Section~\ref{sec:Joint}. Apart from proving the global-optimality of the joint design, tight closed-form approximations are also derived for them to gain additional optimal system-design insights.
	\item Numerical results presented in Section~\ref{sec:results} validate the proposed analysis and provide key insights on the optimal hybrid EBF protocol. The optimized performance variation with critical system parameters is conducted to quantify the achievable gains with respect to perfect CSI based and isotropic transmissions. The relative performance gain achieved by both joint and individual PA and time allocation (TA) schemes is also characterized.
\end{itemize}

Summarizing the notations used in this work, the vectors and matrices are denoted by boldface lowercase and boldface capital letters, respectively. $\mathbf{A}^{\mathrm H}$, $\mathbf{A}^{\mathrm T}$, $\mathbf{A}^{\mathrm *}$, and $\mathbf{A}^{-1}$ respectively denote the Hermitian transpose, transpose, conjugate, and inverse of matrix $\mathbf{A}$. $\mathbf{0}_{n}$, $\mathbf{1}_{n}$,  and $\mathbf{I}_{n}$ respectively  represent the $n\times 1$ zero vector, $n\times 1$  vector with all entries as one, and $n\times n$ identity matrices. With $\mathrm{tr}\left(\mathbf{A}\right)$  being the trace of matrix $\mathbf{A}$ and $[\mathbf{A}]_{i,k}$ denoting its $(i,k)$th element, $[\mathbf{D}]_{i}$ and $[\mathbf{a}]_{i}$ respectively denote the $i$th diagonal entry of the diagonal matrix $\mathbf{D}$ and $i$th entry of the vector $\mathbf{a}$. $\lVert\,\cdot\,\rVert$ and $\left|\,\cdot\,\right|$ respectively represent the Euclidean norm of a complex matrix and the absolute value of a complex scalar. The expectation, covariance, and variance operators have been respectively denoted using $\mathbb{E}\left\lbrace\cdot\right\rbrace$,  $\mathrm{cov}\left\lbrace\cdot\right\rbrace$, and  $\mathrm{var}\left\lbrace\cdot\right\rbrace$. With real and imaginary components of complex quantity $a$ defined using $\mathrm{Re}\left\lbrace a\right\rbrace$ and  $\mathrm{Im}\left\lbrace a\right\rbrace$,  $\phase{\,a}=\mathrm{tan}^{-1}\left(\frac{\mathrm{Im}\left\lbrace a\right\rbrace}{\mathrm{Re}\left\lbrace a\right\rbrace}\right)$ denotes the angle of $a$. Lastly, with $j=\sqrt{-1}$ and $\mathbb{C}$  denoting  complex number set, $\mathbb{C} \mathbb{N}\left(\boldsymbol{\mu},\mathbf{C}\right)$ represents   circularly symmetric complex Gaussian distribution with mean vector $\boldsymbol{\mu}$ and covariance matrix  $\mathbf{C}$. 
 
 
\begin{figure}[!t] 
	\centering\includegraphics[width=3.48in]{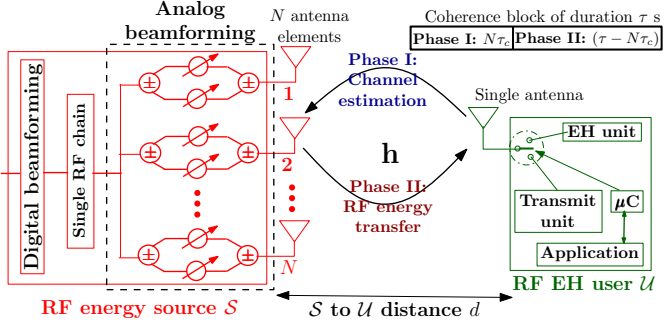}
	\caption{Adopted API model for the DL hybrid EBF from $\mathcal{S}$ using the proposed UL CE via pilot signal transmission by $\mathcal{U}$.} 
	\label{fig:model} 
\end{figure}
 
\section{System Description}\label{sec:model}  
We first present the system model details along with the adopted  channel model and hybrid EBF architecture. Later, we  also discuss the nonlinear RF EH model used in this paper. 
\subsection{Nodes Architecture and MISO Channel Model}\label{sec:arch} 
\textcolor{black}{We consider a MISO wireless RFET  from an $N$ antenna array based RF energy source $\mathcal{S}$ to a single antenna RF EH user $\mathcal{U}$. The detailed system model diagram is presented in Fig.~\ref{fig:model}. The dedicated energy source $\mathcal{S}$ consists of a single RF chain which is shared among the $N\gg1$ antenna elements.} On other end,  EH user $\mathcal{U}$ can be a low-power sensor or IoT device programmed for performing an application-specific operation using its own micro-controller ($\mu C$). We assume $\mathcal{U}$ is solely powered by the energy stored in the EH unit, being replenished via RFET  from $\mathcal{S}$.

With $N\gg 1$, we assume flat quasi-static Rician block fading~\cite[Ch 2.2]{simon2005digital} where the channel impulse response for each communication link remains invariant during a coherence interval of $\tau$ seconds (s) and varies independently across different coherence blocks. The  $\mathcal{S}$-to-$\mathcal{U}$ channel is represented by an $N\times1$ complex vector $\mathbf{h}=\sqrt{\frac{\beta K}{K+1}}\mathbf{h}_{d}+\sqrt{\frac{\beta}{K+1}}\mathbf{h}_s$, where $\mathbf{h}_{d}\in\mathbb{C}^{N\times1}$ is a deterministic complex vector containing the LoS and specular components of the Rician channel vector $\mathbf{h}$, $\beta$ models the large-scale fading between $\mathcal{S}$ and $\mathcal{U}$ which includes both the distance-dependent path loss and shadowing, $K$ is the Rician factor denoting the power ratio between the deterministic and scattered components of the $\mathcal{S}$-to-$\mathcal{U}$ channel. On the other hand,  $\mathbf{h}_{s}\in\mathbb{C}^{N\times1}$ is a complex Gaussian random vector, with independent and identically distributed zero-mean unit-variance entries, representing the scattered components of the $\mathcal{S}$-to-$\mathcal{U}$ channel. So, $\mathbf{h}\sim\mathbb{C} \mathbb{N}\left(\boldsymbol{\mu}_{\mathbf{h}},\mathbf{C}_{\mathbf{h}}\right)$, where $\boldsymbol{\mu}_{\mathbf{h}}=\sqrt{\frac{\beta K}{K+1}}$ $\Big[\sqrt{\alpha_{i_0}}$ $\,\sqrt{\alpha_{1}}\mathrm{e}^{j\theta_{1}\left(\psi\right)}\,\ldots\,\sqrt{\alpha_{N-1}}\mathrm{e}^{j\theta_{N-1}\left(\psi\right)}\Big]^{\mathrm {T}}$ and $\mathbf{C}_{\mathbf{h}}=\frac{\beta}{K+1}\mathbf{I}_N$. Here, $\alpha_{k}$ and $\theta_{k}$ respectively represent the gain of $k$th antenna at $\mathcal{S}$ and its phase shift  with respect to the reference antenna, while $\psi$ is the angle of arrival/departure of the specular component at $\mathcal{S}$ from $\mathcal{U}$. With $\delta$ representing the inter-antenna separation at $\mathcal{S}$, $\theta_{k}\left(\psi\right)\triangleq2\pi k\, \delta\sin\left(\psi\right)$.  

\subsection{Practical Hybrid EBF Architecture under API}\label{sec:API} 
The key idea behind hybrid EBF implementation stems from the result in \cite[Theorem 1]{HBF-Prac1}, where any complex number   $a=\abs{a}\,\mathrm{e}^{\,j\,\phase{\,a}}$ can be alternately represented using a DCPS pair as
\begin{align}\label{eq:complex}
a=\mathrm{e}^{j\left(\cos^{-1}\left(\frac{\abs{a}}{2}\right)+\phase{\,a}\right)}+\mathrm{e}^{-j\left(\cos^{-1}\left(\frac{\abs{a}}{2}\right)-\phase{\,a}\right)}.
\end{align} 
As shown in Fig.~\ref{fig:model}, each antenna element has two DCPSs and one combiner, which in practice suffer from  amplitude and phase errors~\cite{bakr2009impact,PS1,PS2}, that adversely effect the performance of the hybrid EBF. Actually, the latter involves usage of adaptive arrays comprising  multiple antenna elements whose respective beam pattern is shaped by controlling the amplitudes and phases of the RF signals transmitted or received by them.  A precise control over both amplitudes and phases is essential to achieve the desired performance. However, several practical constraints like finite resolution PSs, noise, mismatch in PS circuit elements, and channel uncertainty limit the practically achievable precision~\cite{bakr2009impact,PS1,PS2}. These errors cause an imbalance in the DCPS pair for each antenna element. Some of these error sources are random, and some are fixed which depend on the manufacturing errors and long-term aging effects. These API, which are unpredictable and time varying~\cite{bakr2009impact,PS1,PS2}, can be modeled using random variables with the manufacturing or aging dependent error deciding their means and the random noise based error controlling the variance. Adopting a recently introduced API model~\cite{SPL} for characterizing amplitude and phase errors in practical DCPSs and combiners implementation, the ideal signal in  \eqref{eq:complex}, gets altered to
\begin{align}\label{eq:API-model-complex}
\widetilde{a}=&\,g_{{\mathrm A}_{1}}\,\mathrm{e}^{j\left(\cos^{-1}\left(\frac{\abs{a}}{2}\right)+\phase{\,a}+\phi_{{\mathrm A}_{1}}\right)} +g_{{\mathrm A}_{2}}\,\mathrm{e}^{-j\left(\cos^{-1}\left(\frac{\abs{a}}{2}\right)-\phase{\,a}-\phi_{{\mathrm A}_{2}}\right)},
\end{align}
where $g_{{\mathrm A}_{1}}$ and $g_{{\mathrm A}_{2}}$ respectively represent the amplitude errors due to the API in the first and second DCPS in a pair. Likewise, $\phi_{{\mathrm A}_{1}}$ and  $\phi_{{\mathrm A}_{2}}$ respectively represent the corresponding phase errors.  For the ideal case with no API, $g_{{\mathrm A}_{1}}=g_{{\mathrm A}_{2}}=1$ and $\phi_{{\mathrm A}_{1}}=\phi_{{\mathrm A}_{2}}=0^{\circ}$, which reduces $\widetilde{a}$ in \eqref{eq:API-model-complex} to $a$ in \eqref{eq:complex} . We assume that these random errors in amplitude and phase for each DCPS pair and combiner are independently and uniformly distributed across the different antennas.

\textcolor{black}{Hence, with positive constants $\Delta_{g_i}$ and  $\Delta_{\phi_i}$ representing the errors due to the fixed sources, the amplitude and phase errors, representing the API in practical hybrid architectures, can be respectively modeled as $g_{{\mathrm A}_{i}}$ and $\phi_{{\mathrm A}_{i}}$ defined below
\setcounter{equation}{2}
\begin{equation}\label{eq:g-p-API}
g_{{\mathrm A}_{i}}\triangleq1-\Delta_{g_i}\left(1+\Psi_{g_i}\right),\quad\phi_{{\mathrm A}_{i}}\triangleq \Delta_{\phi_i}\left(1+\Psi_{\phi_i}\right),\;\forall i=1,2,
\end{equation}
where, $\Psi_{g_i}$ and $\Psi_{\phi_i}$, representing the errors due to random sources, follow the uniform distribution with the  respective probability density functions being $f_{\Psi_{g_i}}\left(x\right)= \frac{1}{\Phi_{g_i}},\;\forall x\in\left[-\frac{\Phi_{g_i}}{2},\frac{\Phi_{g_i}}{2}\right]$ and $f_{\Psi_{\phi_i}}\left(x\right)= \frac{1}{\Phi_{\phi_i}},\;\forall x\in\left[-\frac{\Phi_{\phi_i}}{2},\frac{\Phi_{\phi_i}}{2}\right]$. Here, the phase errors are expressed in radians. Uniform distribution is employed because it is commonly adopted for modeling  RF imperfections~\cite{TWC-IQI,bakr2009impact} like PS and oscillator impairments leading to amplitude losses and phase errors due to the usage of low-cost hardware attributing to limited accuracy, and getting influenced by temperature variation  and aging effects.}
Further, under the assumption $0\le \abs{a}\le 2$, which can be easily implemented via the digital beamforming design~\cite{HBF-Prac1}, $\widetilde{a}$ in \eqref{eq:API-model-complex} can be alternatively written as
\begin{align}\label{eq:API-model-complex2}
\widetilde{a}=\Theta\left\lbrace a\right\rbrace\triangleq&\;\mathrm{e}^{j\,\phase{\,a}}\left[\left({\abs{a}}/{2}\right)\left(g_{{\mathrm A}_{1}}\mathrm{e}^{j\phi_{{\mathrm A}_{1}}}+g_{{\mathrm A}_{2}}\mathrm{e}^{j\phi_{{\mathrm A}_{2}}}\right) +\right.\nonumber\\
&\left.j\sqrt{1-\left({\abs{a}}/{2}\right)^2}\left(g_{{\mathrm A}_{1}}\mathrm{e}^{j\phi_{{\mathrm A}_{1}}}-g_{{\mathrm A}_{2}}\mathrm{e}^{j\phi_{{\mathrm A}_{2}}}\right)\right].
\end{align}
We will be using this definition for modeling API in  the analog estimator and precoder designs.

\subsection{Adopted Nonlinear RF Energy Harvesting Model}\label{sec:RFEH}
For the practical RF EH circuits, the harvested direct current (DC) power $ p_h $ is a nonlinear function of the received RF power $ p_r $~\cite{Powercast-P1110,ICC17Wksp,nonRFH2018,Sigmoidal-RFH,RFEH2008} at the input of the RF EH unit performing the RF-to-DC rectification operation. Specifically, $p_h$ depends on the rectification efficiency,   which itself is a function of $ p_r $. Recently~\cite{ICC17Wksp,TWC17}, a piecewise linear approximation (PWLA) was proposed for establishing the relationship between $ p_h $ and $ p_r $ using the function $\mathcal{L}\left\lbrace\cdot\right\rbrace$. \textcolor{black}{Mathematically, considering $L\ge1$ linear pieces, $ p_h =\mathcal{L}\left\lbrace p_r \right\rbrace$ is defined as\enlargethispage*{1\baselineskip}
\begin{align}\label{eq:PWLA0} 
&  p_h=\mathcal{L}\left\lbrace p_r \right\rbrace  \triangleq  \begin{cases}
0, & \text{$ p_r < p_{\mathrm{th}_1},$}\\
\mathcal{A}_i\,  p_r  + \mathcal{B}_i, & \text{$ p_r \in\left[ p_{\mathrm{th}_i}, p_{\mathrm{th}_{i+1}}\right],\,\forall\,i \le L,$}\\
p_{\mathrm{sat}},   & \text{$  p_r > p_{\mathrm{th}_{L+1}}.$}
\end{cases}
\end{align}
Here, $ p_{t}=\left\lbrace  p_{\mathrm{th}_i} \mathrel{}\middle|\mathrel{} 1 \le i \le L+1\right\rbrace\mu$W are thresholds on $ p_r $ defining the boundaries for the $L$ linear pieces with slope $\mathcal{A}=\left\lbrace \mathcal{A}_i \mathrel{}\middle|\mathrel{} 1 \le i \le L\right\rbrace$ and intercept $\mathcal{B}=\left\lbrace \mathcal{B}_i\mathrel{}\middle|\mathrel{} 1 \le i \le L\right\rbrace\mu$W, and constant $p_{\mathrm{sat}}$ is the saturated harvested power for $p_r>p_{\mathrm{th}_{L+1}}.$  In practice for some harvesters, like the Powercast P1110 evaluation board~\cite{Powercast-P1110}, there is a limit on the maximum permissible received RF power  ($p_{\mathrm{th}_{L+1}}=20$dBm) at the input of EH unit to avoid any  damage to the underlying circuit components. Hence, $p_h$ for $p_r>p_{\mathrm{th}_{L+1}}$ is sometimes not defined~\cite[eq. (6)]{ICC17Wksp}.}

\begin{figure}[!t] 
	\centering\includegraphics[width=3.35in]{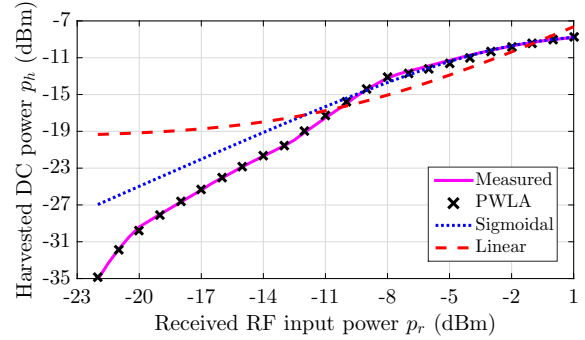}
	\caption{Verifying the quality of PWLA based nonlinear model~\cite{ICC17Wksp} for the RF-EH circuit designed  in~\cite{RFEH2008} for efficient far-field RFET. Other two commonly adopted EH models (sigmoidal~\cite{Sigmoidal-RFH} and linear~\cite{nonRFH2018}) are also plotted.}
	\label{fig:RFEH}
\end{figure}
Using \eqref{eq:PWLA0}, the PWLA for harvested versus received power (HRP) characteristic of the far-field RF EH circuit designed for efficient low-power and long-range RFET  can be obtained with $p_{t}=\{ 6.31,56.23,158.49,$ $562.34,1000,1258.9\}$ $\mu$W as six received threshold powers dividing the HRP characteristic of RF EH circuit designed in~\cite{RFEH2008} into $L=5$ linear pieces having slope $\mathcal{A}=\{0.193,0.375,0.13,0.054,$ $0.028\}$ and intercept $\mathcal{B}=\left\lbrace -0.89,-11.767,30.702,72.372,97.284\right\rbrace \mu$W. \textcolor{black}{As the HRP characteristics are defined only for the input power levels between $-22$dBm to $1$dBm~\cite{RFEH2008}, we set $p_{\mathrm{sat}}=0.25$mW to generate  harvested power results for $p_r>1$dBm $ =1.26$mW.}

In this work we have used this RF EH circuit for investigating the optimized hybrid EBF performance under joint API and CE errors. The goodness of the proposed PWLA model for this practical RF EH  circuit~\cite{RFEH2008} as shown via the log-log plot in Fig.~\ref{fig:RFEH}, is verified by very low  norm of residuals of $2\times10^{-5}$ and root mean square error (RMSE) of $3\times10^{-6}$. In Fig.~\ref{fig:RFEH} we have also plotted the recently proposed sigmoidal (logistic) approximation~\cite{Sigmoidal-RFH} for $ p_h $ as a function of $ p_r $ along with the widely adopted linear fit\footnote{\textcolor{black}{Generally, there are three types of linear EH models~\cite{nonRFH2018}: (a) linear, (b) constant-linear (CL), and (c) constant-linear-constant (CLC). In Fig.~\ref{fig:RFEH}, we plotted the \textit{linear model} which is most commonly adopted due to its analytical simplicity~\cite{Signal-WPT-TC17,CSI-SU-WET,ICASSP18,New-One-Bit,New-TSP,New-WiOpt,HIPT-TWC17}. Whereas, we have considered a more generic PWLA model~\cite{ICC17Wksp} in place of CL (having $L=2$) and CLC (i.e., $L=3$) models.}}. Results show that PWLA provides a much simpler and tighter fit. Therefore, we  use this PWLA in \eqref{eq:PWLA0} for analyzing the RF EH operation.

\section{Least-Squares Hybrid Channel Estimation}\label{sec:CE} 
In this work, we refer the $\mathcal{S}$-to-$\mathcal{U}$ channel as the DL and the $\mathcal{U}$-to-$\mathcal{S}$ link as UL. \textcolor{black}{In contrast to frequency-division duplex (FDD) systems where an estimate of CSI is obtained using feedback schemes, we consider the time-division duplex (TDD) mode of communication in MISO systems~\cite{CSI-SU-WET,ICASSP18,Signal-WPT-TC17}, where the channel reciprocity can be exploited. Hence adopting the TDD mode of communication, where the UL pilot and DL energy  signal transmissions using the same frequency resource are separated in time, the DL channel coefficients can be obtained by estimating them from the UL pilot transmission from $\mathcal{U}$.}  We consider that each coherence interval of $\tau$ seconds (s) is divided into two sub-phases: (a) UL CE phase of $N\tau_c$ s, and (b) DL RFET  phase of $\left(\tau-N\tau_c\right)$ s. \textcolor{black}{During the CE phase,  $\mathcal{U}$ transmits a  continuous-time pilot signal  $\sqrt{2}\,\mathrm{Re}\left\lbrace\mathrm{e}^{-j\,2\pi f_c t}\,\mathrm{s}\left(t\right)\right\rbrace$, having frequency $f_c$  with its baseband representation $\mathrm{s}\left(t\right)=\frac{1}{\sqrt{N\tau_c}}$, satisfying $\int_{0}^{N\tau_c}\abs{\mathrm{s}\left(t\right)}^2\mathrm{d}t=1$.} Thus,  received baseband signal  $\mathbf{y}\in\mathbb{C}^{N\times1}$ at $\mathcal{S}$ is given by
\begin{equation}\label{eq:rxS}
\mathbf{y}\left(t\right)=\sqrt{\mathrm{E}_c}\,\mathbf{h}\,\mathrm{s}\left(t\right)+\mathbf{w}\left(t\right),\quad\forall\,t\in\left[0,N\tau_c\right],
\end{equation} 
where, $\mathrm{E}_c\triangleq p_c\,N\tau_c$ is the energy spent during CE in Joule (J) with $p_c$ denoting the transmit power of $\mathcal{U}$ and $\mathbf{w}\left(t\right)\in\mathbb{C}^{N\times 1}$ is the received complex additive white Gaussian noise (AWGN).

\subsection{Analog Channel Estimator}\label{sec:ana}
Adopting the \textit{antenna-switching} based analog CE approach as proposed in~\cite[Fig. 2]{SPL}, where the parallel estimation of $N$ entries of vector $\mathbf{h}$ over a duration of $N\tau_c$ s reduces to a sequential estimation of each entry  $[\mathbf{h}]_k,\,\forall\,k\in\mathcal{N}$, each over $\tau_c$ s duration. Thus, with $\mathcal{S}$ having only its $k$th antenna active during the $k$th CE sub-phase interval $\tau_{c_k}\triangleq\left(\left(k-1\right)\tau_c,\,k\,\tau_c\right]$, the corresponding analog channel estimator is set as $\mathbf{\bar{f}}_{\mathrm {A_{Id}}}$ defined in~\eqref{eq:fId1}. Under the API at $\mathcal{S}$ as given by \eqref{eq:API-model-complex2} in Section~\ref{sec:API}, the practical entries of the analog channel estimator $\mathbf{f}_{\mathrm{A}}\left(t\right)$ for $t\in\left(\left(k-1\right)\tau_c,\,k\,\tau_c\right],\,\forall\,k\in\mathcal{N}$, which remain the same for each $\tau_c$ duration are~\cite[eq. (7)]{SPL}
\begin{align}\label{eq:fId1}
&[\mathbf{\bar{f}}_{\mathrm A}]_i\triangleq \Theta\left\lbrace[\mathbf{\bar{f}}_{\mathrm {A_{Id}}}]_i\right\rbrace,\; \quad\text{where }\nonumber\\
&[\mathbf{\bar{f}}_{\mathrm {A_{Id}}}]_i=
\begin{cases} 
1,   &   i=k,\\
0,   &   i\neq k,
\end{cases} \qquad\forall\,i\in\mathcal{N}\triangleq\left\lbrace1,2,\ldots,N\right\rbrace.
\end{align} 

\textcolor{black}{Therefore, the entries of analog channel estimator matrix  $\mathbf{F}_{\mathrm{A}}\in\mathbb{C}^{N\times N}$ as set over $N$ sub-phases, with respective intervals   $t\in\left(\left(i-1\right)\tau_c,\,i\,\tau_c\right],\forall i\in\mathcal{N}$, are defined below~\cite{SPL}
\begin{align}\label{eq:ACE-Pb} 
[\mathbf{F}_{\mathrm{A}}]_{ik}=\begin{cases} 
\frac{g_{{\mathrm A}_{i_1}}\mathrm{e}^{j\phi_{{\mathrm A}_{i_1}}} \left(1+j\sqrt{3}\right)\,+\, g_{{\mathrm A}_{i_2}}\mathrm{e}^{j\phi_{{\mathrm A}_{i_2}}}\left(1-j\sqrt{3}\right)}{2},   &   i=k,\\
j\big(g_{{\mathrm A}_{i_1}}\mathrm{e}^{j\phi_{{\mathrm A}_{i_1}}}-g_{{\mathrm A}_{i_2}}\mathrm{e}^{j\phi_{{\mathrm A}_{i_2}}}\big),   &   i\neq k,
\end{cases}
\end{align} 
using the identities, $\mathrm{e}^{j\cos^{-1}\left(\pm\frac{1}{2}\right)}=\frac{1\pm j\sqrt{3}}{2}$ and $\mathrm{e}^{\pm j}=\pm j,$ in \eqref{eq:API-model-complex} for $a=1$ and $a=0$, respectively.}
For ideal (no API)  scenario, $\mathbf{F}_{\mathrm A}=\mathbf{I}_N$  with $g_{{\mathrm A}_{i_k}}=1$ and $\phi_{{\mathrm A}_{i_k}}=0^{\circ},\forall i\in\mathcal{N},k=1,2$. Hence, with this analog channel estimator, the corresponding signal $\mathbf{y}_{\mathrm A}\left(t\right)\in\mathbb{C}^{N\times 1}$ received as an input to digital channel estimator block, and obtained after using \eqref{eq:ACE-Pb} in \eqref{eq:rxS} is given by~\cite[eq. (9)]{SPL}
\begin{align}\label{eq:RX-A}
\left[\mathbf{y}_{\mathrm A}\right]_i\left(t\right)= \sqrt{\mathrm{E}_c}&\, \left[\mathbf{h}_{\mathrm{A}}\right]_i\,\mathrm{s}\left(t\right)+\left[\mathbf{w}_{\mathrm{A}}\right]_i\left(t\right),\nonumber\\
&\forall \left\lbrace t\in\left(\left(i-1\right)\tau_c,\,i\,\tau_c\right]\right\rbrace\wedge\left\lbrace i\in\mathcal{N}\right\rbrace,
\end{align}
where $\mathbf{h}_{\mathrm{A}}\triangleq\mathbf{F}_{\mathrm A}^{\mathrm T}\,\mathbf{h}\in\mathbb{C}^{N\times 1}$ is the effective channel to be estimated and $\mathbf{w}_{\mathrm{A}}\left(t\right)\triangleq\mathbf{F}_{\mathrm A}^{\mathrm T}\, \mathbf{w}\left(t\right)$.  


\subsection{Least-Squares Based Digital Channel Estimator}\label{sec:LSE}
For obtaining the optimal LSE, the received signal $\mathbf{y}(t)$ at the $N$ antennas of $\mathcal{S}$, as defined in \eqref{eq:rxS}, first undergoes the analog CE process as  described by $\mathbf{F}_{\mathrm{A}}$ (cf. \eqref{eq:ACE-Pb}) in Section~\ref{sec:ana}. Thereafter, the match filtering operation is performed to the resulting signal $\mathbf{y}_{\mathrm A}\left(t\right)=\mathbf{F}_{\mathrm A}^{\mathrm T}\,\mathbf{y}\left(t\right)$ by setting the digital channel estimator as $\mathrm{f}_{\mathrm D}\triangleq\frac{\mathrm{s}^*\left(t\right)}{\sqrt{\mathrm{E}_c}}$. Hence, LSE $\widehat{\mathbf{h}}_{\mathrm{A}}$ for the effective channel $\mathbf{h}_{\mathrm{A}}=\mathbf{F}_{\mathrm A}^{\mathrm T}\,\mathbf{h}$ as obtained using \eqref{eq:RX-A} can be derived as shown below
\begin{align}\label{eq:LSE1}
\widehat{\mathbf{h}}_{\mathrm{A}}=&\,\int_{0}^{N\tau_c}\frac{\mathrm{s}^*\left(t\right)}{\sqrt{\mathrm{E}_c}}\,\mathbf{y}_{\mathrm {A}}\left(t\right)\mathrm{d}t= \mathbf{h}_{\mathrm{A}}+\frac{\mathbf{\overline{w}}_{\mathrm{A}}}{\sqrt{\mathrm{E}_c}},
\end{align}
where $\mathbf{\overline{w}}_{\mathrm{A}}=\int_{0}^{N\tau_c}\frac{\mathrm{s}^*\left(t\right)\,\mathbf{F}_{\mathrm A}^{\mathrm T}\,\mathbf{w}\left(t\right)}{\sqrt{\mathrm{E}_c}}\mathrm{d}t$ is the effective AWGN vector influenced by API having zero mean entries. Here, we would like to mention that the proposed LSE $\widehat{\mathbf{h}}_{\mathrm{A}}$ for the effective channel $\mathbf{h}_{\mathrm{A}}$, as obtained using analog and digital channel estimators,  $\mathbf{F}_{\mathrm{A}}$  and  $\mathrm{f}_{\mathrm D}$, yields the \textit{global minimum} value of the objective function $\norm{\widehat{\mathbf{h}}_{\mathrm{A}}-\mathbf{h}_{\mathrm{A}}}^2$ in the conventional LS problem~\cite{kay1993fundamentals}. Hence, $\widehat{\mathbf{h}}_{\mathrm{A}}$ is the global optimal LSE for the effective  $\mathbf{h}_{\mathrm{A}}$ or the actual channel $\mathbf{h}$ under API. Further, these underlying LS estimation errors due to the API and CE uncertainty are independent because their respective sources, i.e., hardware impairments and AWGN, are not related.

\subsection{Optimal Precoder Design}\label{sec:precoder}
The optimal precoder design for the hybrid EBF should be such that it maximizes the received RF power at $\mathcal{U}$ by focusing most of transmit power of $\mathcal{S}$ in the direction of $\mathcal{U}$. Thus, the maximum ratio  transmission (MRT) based precoder design should be selected at $\mathcal{S}$ to maximize the EBF gains. Hence, for implementing the transmit hybrid EBF at $\mathcal{S}$ to maximize the harvested DC power, the digital precoder is set as  $\mathrm{z}_{\mathrm{D}}=\sqrt{p_d}$ and the analog precoder is set as  $\mathbf{z}_{\mathrm{A}}=\frac{\widehat{\mathbf{h}}_{\mathrm{A}}^{\mathrm{H}}}{\norm{\widehat{\mathbf{h}}_{\mathrm{A}}}}\in\mathbb{C}^{1\times N}$. Here, $p_d$ is the transmit power of $\mathcal{S}$ during DL RFET  and $\widehat{\mathbf{h}}_{\mathrm{A}}$ is the LSE for the channel $\mathbf{h}$ as defined in \eqref{eq:LSE1}. However, under API, the analog precoder $\mathbf{z}_{\mathrm{A}}$ gets practically altered to $\overline{\mathbf{z}}_{\mathrm{A}}$
\begin{align}\label{eq:preC}
[\overline{\mathbf{z}}_{\mathrm{A}}]_i\triangleq  \Theta\left\lbrace[\mathbf{z}_{\mathrm{A}}]_i\right\rbrace,\quad\forall\,i\in\mathcal{N}.
\end{align}

\begin{figure}[!t] 
	\centering\includegraphics[width=3in]{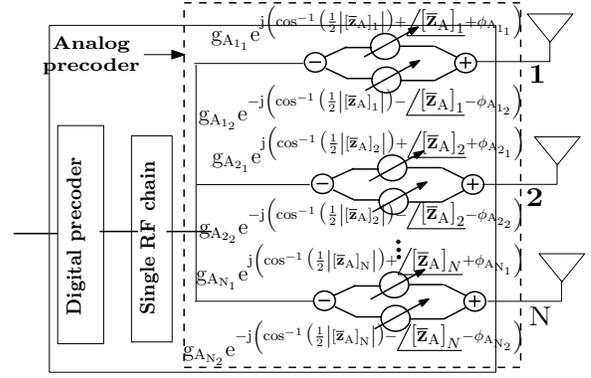}
	\caption{Depicting the alteration to the analog precoder $\mathbf{z}_{\mathrm{A}}$ under practical API in the hybrid EBF implementation.}
	\label{fig:precoder} 
\end{figure}

The entries of this practical analog precoder design  $\overline{\mathbf{z}}_{\mathrm{A}}$ are also depicted in Fig.~\ref{fig:precoder} by showing the equivalent complex baseband signal model for the analog and digital precoder designs. Here, its alternative form is shown, which is reproduced below, was obtained using \eqref{eq:API-model-complex} and \eqref{eq:API-model-complex2}. 
\begin{align}\label{eq:preC2}
[\overline{\mathbf{z}}_{\mathrm{A}}]_i&\,=g_{{\mathrm A}_{i_1}}\,\mathrm{e}^{j\left(\cos^{-1}\left(\frac{\abs{[\overline{\mathbf{z}}_{\mathrm{A}}]_i}}{2}\right)+\phase{[\overline{\mathbf{z}}_{\mathrm{A}}]_i}+\phi_{{\mathrm A}_{i_1}}\right)} +\nonumber\\
&g_{{\mathrm A}_{i_2}}\,\mathrm{e}^{-j\left(\cos^{-1}\left(\frac{\abs{[\overline{\mathbf{z}}_{\mathrm{A}}]_i}}{2}\right)-\phase{[\overline{\mathbf{z}}_{\mathrm{A}}]_i}-\phi_{{\mathrm A}_{i_2}}\right)},\;\forall\,i\in\mathcal{N}.
\end{align}
 
\section{Practically Motivated Tight Approximations}\label{sec:approx}
Here, first in Section~\ref{sec:aprACE}, we present a practically motivated approximation for API-dependent parameters. Then using it, we obtain the statistics of the key parameters based on the LSE $\widehat{\mathbf{h}}_{\mathrm{A}}$. These statistics derived in Section~\ref{sec:stat} and ~\ref{sec:RxPRV-appr}, will be used later in Sections~\ref{sec:analysis} and~\ref{sec:Joint}.

\subsection{Tight Approximation for Analog Channel Estimator Matrix}\label{sec:aprACE} 
In practice, $\Delta_{g_i}\approx\Delta_{\phi_i}\approx\Phi_{g_i}\approx\Phi_{\phi_i}\approx\Delta,\forall i\in\mathcal{N},$ which has very low value, i.e., $\Delta<0.16$. In fact, as also noted in~\cite{PS1,PS2}, for the  practical PSs design, the phase errors are generally much less than $10^{\circ}$. This condition actually on simplification yields $\Delta<0.16$. \textcolor{black}{It may be recalled that this practical range on amplitude and phase errors is also commonly used while investigating the performance under in-phase-and-quadrature-phase-imbalance (IQI) in practical multiantenna systems~\cite{TWC-IQI,bakr2009impact,IET-IQI}. In fact, the in-phase (I) and quadrature-phase (Q) branches used for generating the desired complex signal $a$, play a very similar role as to a DCPS pair and combiner in the hybrid EBF architecture~\cite{HBF-Prac1} as depicted via \eqref{eq:complex}. Furthermore, the random amplitude and phase errors due to IQI are modeled using uniform distribution~\cite[and references therein]{TWC-IQI}, which corroborates our assumption of modeling random API via uniform distribution in Section~\ref{sec:API}. Moreover, since our proposed optimal hybrid CE and EBF protocol holds good for any generic random distribution characterizing API, later  in Section~\ref{sec:results} we have also considered the Gaussian distribution for modeling randomness in API and conducted a performance comparison against the uniform one to gain insights on the impact of different API distributions.}

Following the above discussion, we present a tight approximation for the ACE matrix $\mathbf{F}_{\mathrm{A}}$ by using the practical limits on API. As in practice, for decent quality DCPSs, $\Delta<0.16$. \textcolor{black}{This implies that since $\Delta^2\ll1$, it results in similar practical ranges for the constants $\big(\Delta_{g_i},\Delta_{\phi_i},\Phi_{g_i},\Phi_{\phi_i},\forall i\in$ $\mathcal{N}\big)$ modeling API. In other words,   $\Delta_{g_i}\left(1+\Psi_{g_i}\right)\approx\Delta_{\phi_i}\left(1+\Psi_{\phi_i}\right)\approx\Delta$.  Finally, using it in \eqref{eq:g-p-API} yields the following approximations which hold good for any distribution of $\Psi_{g_i}$ and $\Psi_{\phi_i}$
\begin{equation}\label{eq:API-appr}
g_{{\mathrm A}_{i_k}}\approx 1-\Delta,\qquad
\phi_{{\mathrm A}_{i_k}}\approx \Delta,\qquad
\forall\,i\in\mathcal{N},\;k=1,2,\;
\Delta\ll 1.
\end{equation} }
Applying the above approximation \eqref{eq:API-appr} in API  to \eqref{eq:ACE-Pb} gives: $[\mathbf{F}_{\mathrm{A}}]_{ik}\approx \left(1-\Delta\right)\,\mathrm{e}^{j\Delta},   \,\forall\, i=k\in\mathcal{N}$, and zero for the other entries. From this result along \eqref{eq:API-appr}, it is noted that in practice  the diagonal entries of $\mathbf{F}$ are very close to each other. Whereas, the non-diagonal entries of $\mathbf{F}_{\mathrm{A}}$ are very close to zero. Applying this practically motivated approximation for the API model with $\Delta<0.16$, the following approximation results can be obtained for  matrices involving products of $\mathbf{F}_{\mathrm{A}}$
\begin{equation}\label{eq:approx1}
\mathbf{F}_{\mathrm{A}}^*\mathbf{F}_{\mathrm{A}}^{\mathrm{T}},\,\mathbf{F}_{\mathrm{A}}^{\mathrm{T}}\mathbf{F}_{\mathrm{A}}^*,\,\mathbf{C}_{\mathbf{F}_{\mathrm A}}\approx{N}^{-1}\,\mathrm{tr}\left\lbrace\mathbf{C}_{\mathbf{F}_{\mathrm A}}\right\rbrace \mathbf{I}_N=\sigma^2_{\mathrm{F}_{\mathrm{A}}}\mathbf{I}_N,
\end{equation}
where $\sigma^2_{\mathrm{F}_{\mathrm A}}\triangleq\frac{1}{N}\,\mathrm{tr}\left\lbrace\mathbf{C}_{\mathbf{F}_{\mathrm A}}\right\rbrace$. Thus, all the three API dependent matrices $\mathbf{F}_{\mathrm{A}}^*\mathbf{F}_{\mathrm{A}}^{\mathrm{T}},\,\mathbf{F}_{\mathrm{A}}^{\mathrm{T}}\mathbf{F}_{\mathrm{A}}^*,$ and $\mathbf{C}_{\mathbf{F}_{\mathrm A}}$ can be practically  approximated as the same scaled identity matrix. Next, we use this approximation to derive the distribution for the key API-dependent  parameters.

\subsection{Statistics for LSE-based Key Parameters}\label{sec:stat} 
\subsubsection{Effective AWGN $\mathbf{\overline{w}}_{\mathrm{A}}$}
As discussed above, for practical API with $\Delta<0.16$, the entries of effective AWGN vector $\mathbf{\overline{w}}_{\mathrm{A}}$ are independently and identically distributed (IID) with zero mean entries. Further, the covariance of $\mathbf{\overline{w}}_{\mathrm{A}}$ can be approximated by $\mathbf{C}_{\mathbf{\overline{w}}_{\mathrm {A}}}\triangleq\sigma_{\mathrm w}^{2}\,\mathbf{C}_{\mathbf{F}_{\mathrm A}}$. Here, $\sigma_{\mathrm w}^{2}$ represents the noise power spectral density in Joule (J) and entries of diagonal matrix $\mathbf{C}_{\mathbf{F}_{\mathrm A}}$ are
\begin{align}
[\mathbf{C}_{\mathbf{F}_{\mathrm A}}]_i=&\,\sum_{k=1}^N\abs{[\mathbf{F}_{\mathrm{A}}]_{ik}}^2= g_{{\mathrm A}_{i_1}}^2+g_{{\mathrm A}_{i_2}}^2-g_{{\mathrm A}_{i_1}}g_{{\mathrm A}_{i_2}}\nonumber\\
&\times\Big(\sqrt{3}\sin\left(\phi_{{\mathrm A}_{i_1}}-\phi_{{\mathrm A}_{i_2}}\right)+\cos\left(\phi_{{\mathrm A}_{i_1}}-\phi_{{\mathrm A}_{i_2}}\right)\Big)\nonumber\\
&+\sum\limits_{k=1,k\neq i}^N \bigg(g_{{\mathrm A}_{k_1}}^2+g_{{\mathrm A}_{k_2}}^2-2\,g_{{\mathrm A}_{k_1}}g_{{\mathrm A}_{k_2}}\nonumber\\
&\qquad\quad\times\cos\left(\phi_{{\mathrm A}_{k_1}}-\phi_{{\mathrm A}_{k_2}}\right)\bigg),\quad\forall i\in\mathcal{N}.
\end{align} 
So, with this approximation for practical API, we can rewrite the LSE as defined by \eqref{eq:LSE1} as
\begin{align}\label{eq:LSE2}
\widehat{\mathbf{h}}_{\mathrm{A}}=\mathbf{h}_{\mathrm{A}}+\widetilde{\mathbf{h}}_{\mathrm{A}},
\end{align}
where $\widetilde{\mathbf{h}}_{\mathrm{A}}\sim\mathbb{C} \mathbb{N}\left(\boldsymbol{0}_N,\frac{\sigma_{\mathrm w}^{2}}{\mathrm{E}_c}\,\mathbf{C}_{\mathbf{F}_{\mathrm A}}\right)$ is the LS estimation error~\cite{kay1993fundamentals}, which is a linear function of the effective AWGN vector $\mathbf{\overline{w}}_{\mathrm{A}}$ and independent of the effective channel vector $\mathbf{h}_{\mathrm{A}}$.
 
\subsubsection{Norm $\norm{\widehat{\mathbf{h}}_{\mathrm{A}}}$ of  LSE $\widehat{\mathbf{h}}_{\mathrm{A}}$}\label{sec:normLSE}
The real and imaginary entries of the LSE $\widehat{\mathbf{h}}_{\mathrm{A}}$ follow real and nonzero mean Gaussian distribution, i.e.,   $\mathrm{Re}\left\lbrace[\widehat{\mathrm{h}}_{\mathrm{A}}]_i\right\rbrace\sim\mathbb{N}\left(\mathrm{Re}\left\lbrace\left[\mathbf{F}_{\mathrm A}^{\mathrm T}\,\boldsymbol{\mu}_{\mathbf{h}}\right]_i\right\rbrace,\frac{1}{2}[\mathbf{C}_{\widehat{\mathbf{h}}_{\mathrm{A}}}]_i\right)$ and $\mathrm{Im}\left\lbrace\widehat{\mathrm{h}}_{\mathrm{A_i}}\right\rbrace\sim\mathbb{N}\left(\mathrm{Im}\left\lbrace\left[\mathbf{F}_{\mathrm A}^{\mathrm T}\,\boldsymbol{\mu}_{\mathbf{h}}\right]_i\right\rbrace,\frac{1}{2}[\mathbf{C}_{\widehat{\mathbf{h}}_{\mathrm{A}}}]_i\right)$. Like $\mathbf{\overline{w}}_{\mathrm{A}}$, for practical API, the entries of $\mathbf{h}_{\mathrm{A}}=\mathbf{F}_{\mathrm A}^{\mathrm T}\,\mathbf{h}$,  and hence $\widehat{\mathbf{h}}_{\mathrm{A}}$, are also IID, So, $\mathbf{h}_{\mathrm{A}}\sim\mathbb{C} \mathbb{N}\left(\mathbf{F}_{\mathrm A}^{\mathrm T}\,\boldsymbol{\mu}_{\mathbf{h}},\frac{\beta}{K+1}\,\mathbf{C}_{\mathbf{F}_{\mathrm A}}\right)$.

Thus, the LSE $\widehat{\mathbf{h}}_{\mathrm{A}}\sim\mathbb{C} \mathbb{N}\left(\mathbf{F}_{\mathrm A}^{\mathrm T}\,\boldsymbol{\mu}_{\mathbf{h}},\mathbf{C}_{\widehat{\mathbf{h}}_{\mathrm{A}}}\right)$, where its covariance $\mathbf{C}_{\widehat{\mathbf{h}}_{\mathrm{A}}}\triangleq\mathrm{cov}\left\lbrace \widehat{\mathbf{h}}_{\mathrm{A}} \right\rbrace=\mathbf{F}_{\mathrm{A}}^{\mathrm{T}}\mathbf{C}_{\mathbf{h}}\mathbf{F}_{\mathrm{A}}^*+\frac{\mathbf{C}_{\mathbf{\overline{w}}_{\mathrm {A}}}}{\mathrm{E}_c}$, which can be practically approximated as $\mathbf{C}_{\widehat{\mathbf{h}}_{\mathrm{A}}}\approx\left(\frac{\beta}{K+1}+\frac{\sigma_{\mathrm w}^{2}}{\mathrm{E}_c}\right)\mathbf{C}_{\mathbf{F}_{\mathrm A}}$. So, it includes both the unknown channel state and API information. Using this mentioned distribution $\widehat{\mathbf{h}}_{\mathrm{A}}$ for practical API,  $\frac{2}{\sigma^2_{\widehat{\mathrm{h}}_{\mathrm{A}}}}\,\norm{\widehat{\mathbf{h}}_{\mathrm{A}}}^2$ follows the non-central chi-square distribution with $2N$ degrees of freedom and non-centrality parameter $\frac{2\,\norm{\mathbf{F}_{\mathrm A}^{\mathrm T}\,\boldsymbol{\mu}_{\mathbf{h}}}^2}{\sigma^2_{\widehat{\mathrm{h}}_{\mathrm{A}}}}$. Here, the variance 
$\sigma^2_{\widehat{\mathrm{h}}_{\mathrm{A}}}$ is defined below
\begin{eqnarray}\label{eq:varS-LSE}
\sigma^2_{\widehat{\mathrm{h}}_{\mathrm{A}}}\triangleq\frac{1}{N}\,\mathrm{tr}\left\lbrace\mathbf{C}_{\widehat{\mathbf{h}}_{\mathrm{A}}}\right\rbrace=\left(\frac{\beta}{K+1}+\frac{\sigma_{\mathrm w}^{2}}{\mathrm{E}_c}\right)\sigma^2_{\mathrm{F}_{\mathrm A}}.
\end{eqnarray}
Further, the expectation of $\norm{\widehat{\mathbf{h}}_{\mathrm{A}}}^2$ is given by
\begin{align}\label{eq:exp-Nh}
\mathbb{E}_{\widehat{\mathbf{h}}_{\mathrm{A}}}\left\lbrace\norm{\widehat{\mathbf{h}}_{\mathrm{A}}}^2\right\rbrace&=N\sigma^2_{\widehat{\mathrm{h}}_{\mathrm{A}}}+{\norm{\mathbf{F}_{\mathrm A}^{\mathrm T}\,\boldsymbol{\mu}_{\mathbf{h}}}^2},
\end{align} 
which in turn can be approximated as below after applying  \eqref{eq:approx1} in \eqref{eq:exp-Nh} for practical API-limits
\begin{align}\label{eq:3-apprC}
\mathbb{E}_{\widehat{\mathbf{h}}_{\mathrm{A}}}&\left\lbrace\norm{\widehat{\mathbf{h}}_{\mathrm{A}}}^2\right\rbrace\approx N\sigma^2_{\widehat{\mathrm{h}}_{\mathrm{A}}}+\norm{\boldsymbol{\mu}_{\mathbf{h}}}^2\sigma^2_{\mathrm{F}_{\mathrm{A}}} \nonumber\\
&=\left[N\left(\frac{\beta}{K+1}+\frac{\sigma_{\mathrm w}^{2}}{\mathrm{E}_c}\right)+\norm{\boldsymbol{\mu}_{\mathbf{h}}}^2\right]\sigma^2_{\mathrm{F}_{\mathrm{A}}}.
\end{align} 

We have validated this distribution of $\norm{\widehat{\mathbf{h}}_{\mathrm{A}}}^2$ by verifying the underlying probability density function (PDF) and cumulative distribution function (CDF) via simulations in Section~\ref{sec:results}.

\subsubsection{Conditional distribution $\mathbf{h}\big|\widehat{\mathbf{h}}_{\mathrm{A}}$ of the channel for a given LSE}
Here, we obtain the statistics (mean and variance) for the actual channel $\mathbf{h}$ for the given LSE $\widehat{\mathbf{h}}_{\mathrm{A}}$ for the effective channel  ${\mathbf{h}}_{\mathrm{A}}$. As from \eqref{eq:LSE1}, $\widehat{\mathbf{h}}_{\mathrm{A}}= \mathbf{h}_{\mathrm{A}}+\frac{\mathbf{\overline{w}}_{\mathrm{A}}}{\sqrt{\mathrm{E}_c}}$, the conditional expectation  $\boldsymbol{\mu}_{\mathbf{h}\big|\widehat{\mathbf{h}}_{\mathrm{A}}}\triangleq\mathbb{E}\left\lbrace\mathbf{h}\big|\widehat{\mathbf{h}}_{\mathrm{A}}\right\rbrace$ can be obtained as~\cite{kay1993fundamentals}
\begin{eqnarray}\label{eq:mu0a}
\boldsymbol{\mu}_{\mathbf{h}\big|\widehat{\mathbf{h}}_{\mathrm{A}}}= \mathbb{E}\left\lbrace\mathbf{h}\right\rbrace+\mathrm{cov}\left\lbrace\mathbf{h},\widehat{\mathbf{h}}_{\mathrm{A}}\right\rbrace\mathbf{C}_{\widehat{\mathbf{h}}_{\mathrm{A}}}^{-1}\left(\widehat{\mathbf{h}}_{\mathrm{A}}-\mathbb{E}\left\lbrace\widehat{\mathbf{h}}_{\mathrm{A}}\right\rbrace\right),
\end{eqnarray}
where $\mathrm{cov}\left\lbrace\mathbf{h},\widehat{\mathbf{h}}_{\mathrm{A}}\right\rbrace=\mathbf{C}_{\mathbf{h}}\mathbf{F}_{\mathrm{A}}^*$, $\mathbf{C}_{\mathbf{h}}=\frac{\beta}{K+1},$ and $\mathbf{C}_{\widehat{\mathbf{h}}_{\mathrm{A}}}=\mathbf{F}_{\mathrm{A}}^{\mathrm{T}}\mathbf{C}_{\mathbf{h}}\mathbf{F}_{\mathrm{A}}^*+\frac{\mathbf{C}_{\mathbf{\overline{w}}_{\mathrm {A}}}}{\mathrm{E}_c}$. On using these statistics in \eqref{eq:mu0a} and simplifying, the desired  expectation $\mathbb{E}\left\lbrace\mathbf{h}\big|\widehat{\mathbf{h}}_{\mathrm{A}}\right\rbrace$ is obtained as
\begin{align}\label{eq:mu-exact}
&\boldsymbol{\mu}_{\mathbf{h}\big|\widehat{\mathbf{h}}_{\mathrm{A}}}\triangleq \left(\mathbf{I}_N-\widetilde{\mathbf{C}}\,\mathbf{F}_{\mathrm{A}}^{\mathrm{T}}\right)\boldsymbol{\mu}_{\mathbf{h}}+\widetilde{\mathbf{C}}\,\widehat{\mathbf{h}}_{\mathrm{A}},\quad\text{with }\nonumber\\
&\widetilde{\mathbf{C}}\triangleq\frac{\beta\,\mathbf{F}_{\mathrm A}^*}{K+1} \left(\frac{\beta}{K+1}\,\mathbf{F}_{\mathrm{A}}^{\mathrm{T}}\mathbf{F}_{\mathrm{A}}^*+\frac{\sigma_{\mathrm w}^{2}}{\mathrm{E}_c}\,\mathbf{C}_{\mathbf{F}_{\mathrm A}}\right)^{-1}.
\end{align} 
Similarly, the covariance $\mathbf{C}_{\mathbf{h}\big|\widehat{\mathbf{h}}_{\mathrm{A}}}\triangleq\mathrm{cov}\left\lbrace\mathbf{h}\big|\widehat{\mathbf{h}}_{\mathrm{A}}\right\rbrace$ is given by
\begin{align}\label{eq:cov-exact}
\mathbf{C}_{\mathbf{h}\big|\widehat{\mathbf{h}}_{\mathrm{A}}} &=\mathbf{C}_{\mathbf{h}}-\mathbf{C}_{\mathbf{h}}\mathbf{F}_{\mathrm{A}}^*\left(\frac{\beta\,\mathbf{F}_{\mathrm{A}}^{\mathrm{T}}\mathbf{F}_{\mathrm{A}}^*}{K+1}+\frac{\sigma_{\mathrm w}^{2}\,\mathbf{C}_{\mathbf{F}_{\mathrm A}}}{\mathrm{E}_c}\right)^{-1}\mathbf{F}_{\mathrm{A}}^{\mathrm{T}}\mathbf{C}_{\mathbf{h}}\nonumber\\
&= \left(\mathbf{I}_N-\widetilde{\mathbf{C}}\,\mathbf{F}_{\mathrm{A}}^{\mathrm{T}}\right)\mathbf{C}_{\mathbf{h}}.
\end{align}
On applying the practical API-limit based approximations as defined in \eqref{eq:approx1} to the above result in \eqref{eq:cov-exact}, the following approximation for the covariance of  $\mathbf{h}\big|\widehat{\mathbf{h}}_{\mathrm{A}}$ can be obtained
\begin{align}\label{eq:cov-appr}
\mathbf{C}_{\mathbf{h}\big|\widehat{\mathbf{h}}_{\mathrm{A}}}\approx\frac{\beta\,\sigma_{\mathrm w}^{2}}{\beta\,\mathrm{E}_c+\sigma_{\mathrm w}^{2}\left(K+1\right)}\,\mathbf{I}_N.
\end{align}
Above, we had used \eqref{eq:approx1} for approximating two API-dependent parameters mentioned below
\begin{subequations}
	\begin{eqnarray}\label{eq:apprC}
	&\widetilde{\mathbf{C}}\approx\frac{\beta\,\mathbf{F}_{\mathrm{A}}^*}{\left(K+1\right)\sigma^2_{\widehat{\mathrm{h}}_{\mathrm{A}}}}={\frac{\beta}{K+1}}\left[{\left(\frac{\beta}{K+1}+\frac{\sigma_{\mathrm w}^{2}}{\mathrm{E}_c}\right)\sigma^2_{\mathrm{F}_{\mathrm A}}}\right]^{-1}\mathbf{F}_{\mathrm{A}}^*,
	\end{eqnarray} 
	\begin{align}\label{eq:1-apprC}
	\left(\mathbf{I}_N-\widetilde{\mathbf{C}}\,\mathbf{F}_{\mathrm{A}}^{\mathrm{T}}\right)\approx\frac{\sigma_{\mathrm w}^{2}\left(K+1\right)}{\beta\,\mathrm{E}_c+\sigma_{\mathrm w}^{2}\left(K+1\right)}\textcolor{black}{\;\mathbf{I}_N}.
	\end{align}
\end{subequations}

\subsection{Received energy signal $\Upsilon_{\mathrm{h}}$ at $\mathcal{U}$ during DL RFET  phase}\label{sec:RxPRV-appr}
As discussed in Section~\ref{sec:precoder}, the analog precoder design set to $\mathbf{z}_{\mathrm{A}}=\frac{\widehat{\mathbf{h}}_{\mathrm{A}}^{\mathrm{H}}}{\norm{\widehat{\mathbf{h}}_{\mathrm{A}}}}\in\mathbb{C}^{1\times N}$. \textcolor{black}{Though this precoder $\mathbf{z}_{\mathrm{A}}$ actually gets altered to $\overline{\mathbf{z}}_{\mathrm{A}}$ defined in \eqref{eq:preC} due to the underlying API, which are not known and thus are not compensated, we have used $\mathbf{z}_{\mathrm{A}}$ for the theoretical investigation in Section~\ref{sec:analysis} and  optimization in  Section~\ref{sec:Joint}.}  Consequently, the corresponding random RF energy signal, as received at $\mathcal{U}$ during the DL RFET  phase, is given by $\Upsilon_{\mathrm{h}}\triangleq\frac{\widehat{\mathbf{h}}_{\mathrm{A}}^\mathrm{H}\,\mathbf{h}}{\norm{\widehat{\mathbf{h}}_{\mathrm{A}}}}$. This random variable will be used  for investigating the optimal resource allocation to maximize the harvested energy performance in Section~\ref{sec:Joint}. Below, we derive the distribution of $\Upsilon_{\mathrm{h}}$ for a given LSE $\widehat{\mathbf{h}}_{\mathrm{A}}$.

\begin{lemma}\label{lem:gam}\itshape
	Given the LSE $\widehat{\mathbf{h}}_{\mathrm{A}}$, $\frac{1}{{\sigma}^2_{\Upsilon_{\mathrm{h}}|\widehat{\mathbf{h}}_{\mathrm{A}}}}\abs{\Upsilon_{\mathrm{h}}}^2$ follows non-central chi-square distribution with two degrees of freedom and the non-centrality parameter $\frac{\norm{\boldsymbol{\mu}_{\Upsilon_{\mathrm{h}}|\widehat{\mathbf{h}}_{\mathrm{A}}}}^2}{{\sigma}^2_{\Upsilon_{\mathrm{h}}|\widehat{\mathbf{h}}_{\mathrm{A}}}}$. Here, the statistics  ${\mu}_{\Upsilon_{\mathrm{h}}|\widehat{\mathbf{h}}_{\mathrm{A}}}$ and ${\sigma}^2_{\Upsilon_{\mathrm{h}}\big|\widehat{\mathbf{h}}_{\mathrm{A}}}$ denote the mean and variance  of the conditional random variable $\Upsilon_{\mathrm{h}}\big|\widehat{\mathbf{h}}_{\mathrm{A}}$,  respectively.
\end{lemma}
\begin{IEEEproof}	
	For the given LSE $\widehat{\mathbf{h}}_{\mathrm{A}}$,  $\Upsilon_{\mathrm{h}}=\frac{\widehat{\mathbf{h}}_{\mathrm{A}}^\mathrm{H}}{\norm{\widehat{\mathbf{h}}_{\mathrm{A}}}}\,\mathbf{h}$, follows the same distribution as $\mathbf{h}$, i.e., nonzero complex Gaussian. Below, we obtain the required statistics to obtain the distribution, i.e., mean $\mathbb{E}\left\lbrace\Upsilon_{\mathrm{h}}\big|\widehat{\mathbf{h}}_{\mathrm{A}}\right\rbrace$ and variance  $\mathrm{var}\left\lbrace\Upsilon_{\mathrm{h}}\big|\widehat{\mathbf{h}}_{\mathrm{A}}\right\rbrace$ of $\Upsilon_{\mathrm{h}}\big|\widehat{\mathbf{h}}_{\mathrm{A}}$.
	\begin{subequations}
		\begin{align}\label{eq:muL}
		{\mu}_{\Upsilon_{\mathrm{h}}\big|\widehat{\mathbf{h}}_{\mathrm{A}}}\triangleq&\,\mathbb{E}\left\lbrace\Upsilon_{\mathrm{h}}\big|\widehat{\mathbf{h}}_{\mathrm{A}}\right\rbrace=\frac{\widehat{\mathbf{h}}_{\mathrm{A}}^\mathrm{H}}{\norm{\widehat{\mathbf{h}}_{\mathrm{A}}}}\,\mathbb{E}\left\lbrace\mathbf{h}\big|\widehat{\mathbf{h}}_{\mathrm{A}}\right\rbrace\nonumber\\
		=&\, \frac{\widehat{\mathbf{h}}_{\mathrm{A}}^\mathrm{H}}{\norm{\widehat{\mathbf{h}}_{\mathrm{A}}}}\left[\left(\mathbf{I}_N-\widetilde{\mathbf{C}}\,\mathbf{F}_{\mathrm{A}}^{\mathrm{T}}\right)\boldsymbol{\mu}_{\mathbf{h}}+\widetilde{\mathbf{C}}\,\widehat{\mathbf{h}}_{\mathrm{A}}\right],
		\end{align}			
		\begin{align}\label{eq:covL}
		{\sigma}^2_{\Upsilon_{\mathrm{h}}\big|\widehat{\mathbf{h}}_{\mathrm{A}}}\triangleq&\,\mathrm{var}\left\lbrace\Upsilon_{\mathrm{h}}\big|\widehat{\mathbf{h}}_{\mathrm{A}}\right\rbrace=\frac{\widehat{\mathbf{h}}_{\mathrm{A}}^\mathrm{H}\;\mathrm{cov}\left\lbrace\mathbf{h}\big|\widehat{\mathbf{h}}_{\mathrm{A}}\right\rbrace\,\widehat{\mathbf{h}}_{\mathrm{A}}}{\norm{\widehat{\mathbf{h}}_{\mathrm{A}}}^2}\nonumber\\ 
		\approx&\,\frac{\beta\,\sigma_{\mathrm w}^{2}}{\beta\,\mathrm{E}_c+\sigma_{\mathrm w}^{2}\left(K+1\right)},
		\end{align}
	\end{subequations}
	where \eqref{eq:muL} is obtained using \eqref{eq:mu-exact}, and \eqref{eq:covL} after using approximation \eqref{eq:cov-appr} in  \eqref{eq:cov-exact}. Hence, for the given LSE $\widehat{\mathbf{h}}_{\mathrm{A}}$, 
	$\Upsilon_{\mathrm{h}}\sim\mathbb{C} \mathbb{N}\left({\mu}_{\Upsilon_{\mathrm{h}}\big|\widehat{\mathbf{h}}_{\mathrm{A}}},\,{\sigma}^2_{\Upsilon_{\mathrm{h}}\big|\widehat{\mathbf{h}}_{\mathrm{A}}}\right)$. Hence, the normalized random variable $\frac{\abs{\Upsilon_{\mathrm{h}}}^2}{{\sigma}^2_{\Upsilon_{\mathrm{h}}|\widehat{\mathbf{h}}_{\mathrm{A}}}}$ follows the mentioned non-central chi-square distribution. Further, on using   approximations \eqref{eq:apprC} and \eqref{eq:1-apprC} defined for practical API settings:\\ $\frac{1}{{\sigma}^2_{\Upsilon_{\mathrm{h}}|\widehat{\mathbf{h}}_{\mathrm{A}}}}\mathbb{E}_{\widehat{\mathbf{h}}_{\mathrm{A}}}\left\lbrace\norm{\boldsymbol{\mu}_{\Upsilon_{\mathrm{h}}\big|\widehat{\mathbf{h}}_{\mathrm{A}}}}^2\right\rbrace\approx\frac{\norm{\boldsymbol{\mu}_{\mathbf{h}}}^2}{{\sigma}^2_{\Upsilon_{\mathrm{h}}|\widehat{\mathbf{h}}_{\mathrm{A}}}}+\frac{\beta\,\mathrm{E}_c\,N}{\left(K+1\right)\sigma_{\mathrm w}^{2}}$.
\end{IEEEproof}	
\medskip
 
\section{Average Harvested Energy due to Hybrid EBF}\label{sec:analysis}
In this section we derive the expression for the average harvested  energy at the EH $\mathcal{U}$ due to MRT from $\mathcal{S}$ during the DL RFET using the LSE $\widehat{\mathbf{h}}_{\mathrm{A}}$. In this regard we first revisit some basics of received power analysis over Rician channels. Thereafter, we discourse the optimal transmit hybrid EBF at $\mathcal{S}$ based on the proposed LSE as presented in Section~\ref{sec:CE}. Lastly, since in general the harvested energy cannot be expressed in closed form, we present a practically-motivated tight analytical approximation for it by using the developments in Sections~\ref{sec:stat} and~\ref{sec:RxPRV-appr}.

\subsection{Exact Average Harvested DC Power Analysis}\label{sec:EH}
Following the result outlined in Lemma~\ref{lem:RV-RxRFP}, we note that the received RF power $ p_r \triangleq  p_d \left|\Upsilon_{\mathrm{h}}\right|^2$ at $\mathcal{U}$ due to DL RFET  from $\mathcal{S}$ follows non-central chi-square distribution with two degrees of freedom, Rice factor $K$ and mean  $\mu_{p_r}$. Thus, the PDF  of the received power $ p_r $ for $x\ge0$ is given by~\cite{simon2005digital}
\begin{align}\label{eq:PDF}
f_{ p_r }\left(x,K,\mu_{p_r}\right)=\frac{\mathrm{e}^{-\frac{(K+1) x}{\mu_{p_r}}-K}}{\mu_{p_r}(K+1)^{-1}} \;\mathbf{I_0}\left(2 \sqrt{\frac{K (K+1) x}{\mu_{p_r}}}\right),
\end{align}
where $\mathbf{I_m}\left(\cdot\right)$ is the modified Bessel function of first kind with order $\mathbf{m}$.  Further, CDF $F_{ p_r }\left(x\right)=\mathrm{Pr}\left\lbrace p_r \le x\right\rbrace$  of $ p_r $ is:
\begin{equation}\label{eq:CDF}
F_{ p_r }\left(x\right)=1-Q_1\left(\sqrt{2 K},\sqrt{\frac{2(K+1) x}{\mu_{p_r}}}\right),\quad x\ge0,
\end{equation}
where $Q_1\left(\cdot,\cdot\right)$ is the first order Marcum Q-function~\cite{integrals}.

Using the relationship $ p_h =\mathcal{L}\left\lbrace p_r \right\rbrace$ from \eqref{eq:PWLA0} along with PDF $f_{ p_r }$ and CDF $F_{ p_r }$ of received power $ p_r $ defined in \eqref{eq:PDF} and \eqref{eq:CDF},  PDF of harvested power $ p_h $ for $x\ge0$ is given by
\begin{align}\label{eq:PWLA2}
f_{ p_h }\left(x\right) \triangleq  \begin{cases}
		\frac{\frac{1}{\mathcal{A}_j}\,f_{ p_r }\left(\frac{x-\mathcal{B}_j}{\mathcal{A}_j},K,\mu_{p_r}\right)}{F_{ p_r }\left( p_{\mathrm{th}_{L+1}}\right)-F_{ p_r }\left( p_{\mathrm{th}_1}\right)}, & \text{$ p_{\mathrm{th}_j}\le\frac{x-\mathcal{B}_j}{\mathcal{A}_j}\le p_{\mathrm{th}_{j+1}},$}\\&\text{$\forall j\in 1,2,\ldots L$,}\\
		0,   & \text{otherwise.}
		\end{cases} 
\end{align}  
Thus, using \eqref{eq:PWLA2}, the mean harvested DC power    is given by $\mu_{ p_h }\triangleq\mathbb{E}\left\lbrace p_h \right\rbrace=\int_0^{\infty}x\; f_{ p_h }\left(x\right)\mathrm{d}x.$ 
Although, it is difficult to obtain a closed-form expression for $\mu_{ p_h }$, an alternate representation in the form of an infinite series was derived in \cite[eq. (8)]{TWC17}. However, for analytical tractability, we use a simpler representation in the form of a tight approximation based on the Jensen's inequality~\cite{boyd}, i.e., $\mu_{ p_h }=\mathbb{E}\left\lbrace\mathcal{L}\left\lbrace p_r \right\rbrace\right\rbrace\le \mathcal{L}\left\lbrace\mathbb{E}\left\lbrace p_r \right\rbrace\right\rbrace=\mathcal{L}\left\lbrace\mu_{p_r}\right\rbrace$, which is defined as below
\begin{align}\label{eq:PWLA1}
\mu_{ p_h }&=\,\mathbb{E}\left\lbrace\mathcal{L}\left\lbrace p_r \right\rbrace\right\rbrace\approx\,\widehat{\mu}_{ p_h}\nonumber\\
&\triangleq  \begin{cases}
0, & \text{$\mu_{p_r}< p_{\mathrm{th}_1},$}\\
\mathcal{A}_i\; \mu_{p_r} + \mathcal{B}_i, & \text{$\mu_{p_r}\in\left[ p_{\mathrm{th}_i}, p_{\mathrm{th}_{i+1}}\right],\; \forall i \le L,$}\\
\text{Not applicable},   & \text{$ \mu_{p_r}> p_{\mathrm{th}_{L+1}}.$}
\end{cases}
\end{align}

\subsection{Transmit Hybrid RF Energy Beamforming}\label{sec:TxEBF}
As mentioned earlier in Section~\ref{sec:precoder}, for implementing the transmit hybrid EBF at $\mathcal{S}$ to maximize the harvested DC power at $\mathcal{U}$, the digital precoder is set as  $\mathrm{z}_{\mathrm{D}}=\sqrt{p_d}$ and analog precoder as  $\mathbf{z}_{\mathrm{A}}=\frac{\widehat{\mathbf{h}}_{\mathrm{A}}^{\mathrm{H}}}{\norm{\widehat{\mathbf{h}}_{\mathrm{A}}}}$. \textcolor{black}{From Section~\ref{sec:RxPRV-appr}, we recall that since API estimation and compensation is not investigated in this work, we have used  $\mathbf{z}_{\mathrm{A}}$, instead of $\overline{\mathbf{z}}_{\mathrm{A}}$, as the analog precoder  for the underlying investigation based on the LSE $\widehat{\mathbf{h}}_{\mathrm{A}}$ for the effective channel $\mathbf{h}_{\mathrm{A}}=\mathbf{F}_{\mathrm A}^{\mathrm T}\,\mathbf{h}$. Further, since API are slowly varying processes because the factors influencing them like aging, hardware  temperature variation, and manufacturing impairments change slowly with time, API mitigation can be incorporated via  calibration methods relying on mutual coupling between  antenna elements~\cite{Myths}. However, the detailed API compensation is out of the scope of this work.}

\textcolor{black}{So, ignoring API compensation, the MRT-based  precoding $\mathbf{z}_{\mathrm{A}}$  enables that the signals emitted from different antennas add coherently at the EH user $\mathcal{U}$. $\mathcal{S}$ transmits a continuous energy signal
$\sqrt{2}\,\mathrm{Re}\left\lbrace\mathrm{e}^{-j\,2\pi f_c t}\,\mathrm{s}_d\left(t\right)\right\rbrace$ to $\mathcal{U}$, satisfying $\int_{0}^{\tau-N\tau_c}\abs{\mathrm{s}_d\left(t\right)}^2\mathrm{d}t={\tau-N\tau_c}$.} Hence, the transmit signal $\mathbf{x}_{_\mathcal{S}}\left(t\right)=\mathrm{z}_{\mathrm{D}}\,\mathbf{z}_{\mathrm{A}}\,\mathrm{s}_d\left(t\right)$ and the received baseband signal $\mathrm{y}_{_\mathcal{U}}\left(t\right)$ at the EH user  $\mathcal{U}$ is given by
\begin{equation}\label{eq:rxU2}
\mathrm{y}_{_\mathcal{U}}\left(t\right)=\sqrt{\mathrm{E}_d}\;\mathbf{x}_{_\mathcal{S}}\left(t\right)\,\mathbf{h}+\mathrm{w}_{_\mathcal{U}}\left(t\right),\quad\forall\,t\in\left[0,\tau-N\tau_c\right],
\end{equation} 
where $\mathrm{E}_d=\left(\tau-N\tau_c\right)p_d$ is the downlink array transmit energy expenditure at $\mathcal{S}$ in J and $\mathrm{w}_{_\mathcal{U}}\left(t\right)$ is the AWGN received at $\mathcal{U}$. So, the received RF energy $\mathrm{E}_r$ in J is given by
\begin{equation}
\mathrm{E}_r=\left(\tau-N\tau_c\right)\,p_r=\left(\tau-N\tau_c\right)\, p_d\abs{\Upsilon_{\mathrm{h}}}^2,
\end{equation}
where $p_r\triangleq p_d\abs{\Upsilon_{\mathrm{h}}}^2$ is the received RF power at $\mathcal{U}$. 
Hence, the resulting energy harvested $\mathrm{E}_h$ in J is given by $
\mathrm{E}_h\triangleq\left(\tau-N\tau_c\right)\mathcal{L}\left\lbrace p_r\right\rbrace=\left(\tau-N\tau_c\right)\mathcal{L}\left\lbrace p_d\abs{\Upsilon_{\mathrm{h}}}^2\right\rbrace.$  Here, $\mathrm{E}_h$ is a random variable because  $\abs{\Upsilon_{\mathrm{h}}}^2$ as mentioned in Lemma~\ref{lem:gam} follows the non-central chi-square distribution. Thus, the mean harvested energy $\mu_{\mathrm{E}_h}\triangleq\mathbb{E}\left\lbrace \mathrm{E}_h\right\rbrace$ can be obtained in terms of the mean harvested power $\mu_{ p_h }=\mathbb{E}\left\lbrace \mathcal{L}\left\lbrace p_r\right\rbrace\right\rbrace$, and its definition in integral form using \eqref{eq:PWLA2} is given below
\begin{equation}\label{eq:Eh-exact}
\mu_{\mathrm{E}_h}\triangleq\left(\tau-N\tau_c\right)\mu_{ p_h }=\left(\tau-N\tau_c\right) \displaystyle\int_0^{\infty}x\; f_{ p_h }\left(x\right)\mathrm{d}x.
\end{equation}
However, for analytical tractability of the above integral of $\mu_{\mathrm{E}_h}$, we use a simpler representation in the form of a tight approximation $\widehat\mu_{\mathrm{E}_h}$ based on the Jensen's inequality~\cite{boyd}, i.e., using \eqref{eq:PWLA1}
\begin{align}\label{eq:Es}
\mu_{\mathrm{E}_h}\approx\widehat\mu_{\mathrm{E}_h}\triangleq\left(\tau-N\tau_c\right)\left[\mathcal{A}_{i_0}\,\mu_{p_r}+\mathcal{B}_{i_0}\right],
\end{align}
where ${i_0}\triangleq\left\lbrace  i\mathrel{}\middle|\mathrel{}\mu_{p_r}\in\left[p_{\mathrm{th}_i},p_{\mathrm{th}_{i+1}}\right],\,1\le i\le L\right\rbrace$.  Next, we outline a key result on $\mu_{p_r}=\mathbb{E}\left\lbrace p_r\right\rbrace$. 
\begin{lemma}\itshape
	Using the conditional statistics \eqref{eq:mu-exact} and \eqref{eq:cov-exact}, the average received power at $\mathcal{U}$ is
	\begin{align}\label{eq:mean-RxP}
	&\mu_{p_r}=\, p_d\,\mathbb{E}\left\lbrace\left|\frac{\widehat{\mathbf{h}}_{\mathrm{A}}^\mathrm{H}\,\mathbf{h}}{\norm{\widehat{\mathbf{h}}_{\mathrm{A}}}}\right|^2\right\rbrace\nonumber\\
	&=\,p_d\, \mathbb{E}_{\widehat{\mathbf{h}}_{\mathrm{A}}} \left\lbrace\frac{\widehat{\mathbf{h}}_{\mathrm{A}}^{\mathrm{H}}}{\norm{\widehat{\mathbf{h}}_{\mathrm{A}}}^2}\left(\mathbf{C}_{\mathbf{h}\big|\widehat{\mathbf{h}}_{\mathrm{A}}}+\boldsymbol{\mu}_{\mathbf{h}\big|\widehat{\mathbf{h}}_{\mathrm{A}}} \boldsymbol{\mu}_{\mathbf{h}\big|\widehat{\mathbf{h}}_{\mathrm{A}}}^{\mathrm{H}} \right)\widehat{\mathbf{h}}_{\mathrm{A}}\right\rbrace.
	\end{align}
\end{lemma}
\begin{IEEEproof}
	Following   Section~\ref{sec:EH}, the average received RF  power $\mu_{p_r}$ at $\mathcal{U}$ can be derived as
	\begin{align}\label{eq:mean-RxP0}
	\mu_{p_r}&=\, p_d\,\mathbb{E}_{\widehat{\mathbf{h}}_{\mathrm{A}}} \left\lbrace\frac{\mathbb{E}_{\mathbf{h}\big|\widehat{\mathbf{h}}_{\mathrm{A}}}\left\lbrace\left(\widehat{\mathbf{h}}_{\mathrm{A}}^\mathrm{H}\,\mathbf{h}\right)\left(\widehat{\mathbf{h}}_{\mathrm{A}}^\mathrm{H}\,\mathbf{h}\right)^{\mathrm{H}}\right\rbrace}{\norm{\widehat{\mathbf{h}}_{\mathrm{A}}}^2}\right\rbrace\nonumber\\
	&=\,p_d\,\mathbb{E}_{\widehat{\mathbf{h}}_{\mathrm{A}}} \left\lbrace\frac{\widehat{\mathbf{h}}_{\mathrm{A}}^{\mathrm{H}}\;\mathbb{E}\left\lbrace\mathbf{h}\,\mathbf{h}^\mathrm{H}\big|\widehat{\mathbf{h}}_{\mathrm{A}}\right\rbrace\widehat{\mathbf{h}}_{\mathrm{A}}}{\norm{\widehat{\mathbf{h}}_{\mathrm{A}}}^2}\right\rbrace.
	\end{align}
	Using the definition in \eqref{eq:hh} along with \eqref{eq:mu-exact} and \eqref{eq:cov-exact}, yields the desired expression for $\mu_{p_r}$.
	\begin{align}\label{eq:hh}
	\mathbb{E}\left\lbrace\mathbf{h}\,\mathbf{h}^\mathrm{H}\big|\widehat{\mathbf{h}}_{\mathrm{A}}\right\rbrace=&\,\mathbf{C}_{\mathbf{h}\big|\widehat{\mathbf{h}}_{\mathrm{A}}}+\boldsymbol{\mu}_{\mathbf{h}\big|\widehat{\mathbf{h}}_{\mathrm{A}}}\,\boldsymbol{\mu}_{\mathbf{h}\big|\widehat{\mathbf{h}}_{\mathrm{A}}}^{\mathrm{H}}. 
	\end{align}
	This completes the proof after some  simplifications.
\end{IEEEproof}

\begin{corollary}\itshape
	The mean received power with perfect CSI availability and no API is given by	  
	\begin{align}\label{eq:ideal}
	\mu_{p_r,\mathrm{id}}\triangleq&\, p_d\,\mathbb{E}\left\lbrace\left|\frac{\mathbf{h}^\mathrm{H}\,\mathbf{h}}{\norm{\mathbf{h}}}\right|^2\right\rbrace  
	= p_d\left[\norm{\boldsymbol{\mu}_{\mathbf{h}}}^2+\frac{\beta\,N}{K+1}\right].
	\end{align}
	
	On the other hand, for the isotropic transmission~\cite[Ch 2.2]{antenna} with $\mathbf{z}_{\mathrm{A}}=\mathbf{1}_{N}^{\mathrm{H}}$, the mean received RF power $\mu_{p_r,\mathrm{iso}}$ is given by
	\begin{align}\label{eq:iso}
	\mu_{p_r,\mathrm{iso}}\triangleq&\, p_d\,\mathbb{E}\left\lbrace\left|\frac{\mathbf{1}_{N}^\mathrm{H}\,\mathbf{h}}{\norm{\mathbf{1}_{N}}}\right|^2\right\rbrace = p_d\,\mathbb{E}\left\lbrace\left|\frac{1}{\sqrt{N}}\sum_{i=1}^N\,[\mathbf{h}]_i\right|^2\right\rbrace \nonumber\\
	=&\, p_d\left[\frac{1}{N}\norm{\mathbf{1}_{N}^\mathrm{H}\,\boldsymbol{\mu}_{\mathbf{h}}}^2+\frac{\beta}{K+1}\right].
	\end{align}
\end{corollary}

\begin{corollary}\itshape
	The average received powers $\mu_{p_r,\mathrm{id}}^{\mathrm{ray}}$ and $
	\mu_{p_r,\mathrm{iso}}^{\mathrm{ray}}$ for the ideal case with perfect CSI and isotropic transmission over Rayleigh fading channels are respectively defined as 
	\begin{equation}
	\mu_{p_r,\mathrm{id}}^{\mathrm{ray}}\triangleq p_d\,\beta\,N,\quad\text{ and}\quad
	\mu_{p_r,\mathrm{iso}}^{\mathrm{ray}}\triangleq p_d\, \beta.
	\end{equation} 
\end{corollary}
\begin{IEEEproof}
	This can be easily obtained after substituting $K=0$ and $\boldsymbol{\mu}_{\mathbf{h}}=\mathbf{0}_N$ (Rayleigh fading properties) in the results \eqref{eq:ideal} and \eqref{eq:iso} as defined for the Rician fading channels.
\end{IEEEproof}

\subsection{Tight Closed-Form Approximation for the Average Received RF Power at the EH User} 
Using the approximations defined in Sections~\ref{sec:aprACE} and~\ref{sec:stat} for the practical limits on API, we next obtain a tight analytical approximation for the mean received RF power defined in \eqref{eq:mean-RxP}.

\begin{lemma}\label{lem:RV-RxRFP}\itshape
	A tight analytical approximation $\widehat{\mu}_{p_r}$ for the mean received RF power $\mu_{p_r}$ at $\mathcal{U}$ using the practical values for the parameter characterizing API, i.e., $\Delta<0.16$, is given by
	\begin{align}\label{eq:mean-aprRxP}
	\widehat{\mu}_{p_r}\triangleq&\,p_d \left[\norm{\boldsymbol{\mu}_{\mathbf{h}}}^2+\frac{\beta\,N}{K+1}-\frac{\beta\,\sigma_{\mathrm w}^{2}\left(N-1\right)}{\beta N p_c\tau_c+\sigma_{\mathrm w}^{2}\left(K+1\right)}\right].
	\end{align}
\end{lemma}
\begin{IEEEproof}	
	From the approximation \eqref{eq:1-apprC} in \eqref{eq:mu-exact}, we obtain
	\begin{align}\label{eq:mu-intr1}
	\boldsymbol{\mu}_{\mathbf{h}\big|\widehat{\mathbf{h}}_{\mathrm{A}}}\approx\frac{\sigma_{\mathrm w}^{2}\left(K+1\right)}{\beta\,\mathrm{E}_c+\sigma_{\mathrm w}^{2}\left(K+1\right)}\,\boldsymbol{\mu}_{\mathbf{h}}+\widetilde{\mathbf{C}}\,\widehat{\mathbf{h}}_{\mathrm{A}}.
	\end{align}		
	Using this approximation  for $\boldsymbol{\mu}_{\mathbf{h}\big|\widehat{\mathbf{h}}_{\mathrm{A}}}$ in \eqref{eq:mu-intr1} along with the ones presented in \eqref{eq:apprC} and \eqref{eq:1-apprC}, we can derive the following result for the expectation $\mathbb{E}_{\widehat{\mathbf{h}}_{\mathrm{A}}}\left\lbrace\boldsymbol{\mu}_{\mathbf{h}\big|\widehat{\mathbf{h}}_{\mathrm{A}}}\,\boldsymbol{\mu}_{\mathbf{h}\big|\widehat{\mathbf{h}}_{\mathrm{A}}}^{\mathrm{H}}\right\rbrace$.
	\begin{align}\label{eq:2-apprC}
	\mathbb{E}_{\widehat{\mathbf{h}}_{\mathrm{A}}}&\left\lbrace\boldsymbol{\mu}_{\mathbf{h}\big|\widehat{\mathbf{h}}_{\mathrm{A}}}\,\boldsymbol{\mu}_{\mathbf{h}\big|\widehat{\mathbf{h}}_{\mathrm{A}}}^{\mathrm{H}}\right\rbrace \approx\Bigg[\frac{\left(\beta\,\mathrm{E}_c\right)^2}{\sigma^2_{\mathrm{F}_{\mathrm{A}}}}\,\mathbb{E}_{\widehat{\mathbf{h}}_{\mathrm{A}}}\left\lbrace\norm{\widehat{\mathbf{h}}_{\mathrm{A}}}^2\right\rbrace+\nonumber\\
		&\qquad\quad\left[\left(\sigma_{\mathrm w}^{2}\left(K+1\right)\right)^2+2\,\sigma_{\mathrm w}^{2}\left(K+1\right)\beta\,\mathrm{E}_c\right]\norm{\boldsymbol{\mu}_{\mathbf{h}}}^2\Bigg]\nonumber\\
		&\qquad\qquad\times{\left(\beta\,\mathrm{E}_c+\sigma_{\mathrm w}^{2}\left(K+1\right)\right)^{-2}}\;\mathbf{I}_N\nonumber\\
	&=\left[\norm{\boldsymbol{\mu}_{\mathbf{h}}}^2+\frac{\beta^2\,\mathrm{E}_c\,N}{\left(K+1\right)\left(\beta\,\mathrm{E}_c+\sigma_{\mathrm w}^{2}\left(K+1\right)\right)}\right]\mathbf{I}_N,
	\end{align}
	where \eqref{eq:2-apprC} is obtained on substituting the approximation~\eqref{eq:3-apprC}. Next, on applying these developments \eqref{eq:cov-appr} and \eqref{eq:2-apprC} along with the linearity of expectation property in \eqref{eq:mean-RxP}, we obtain
	\begin{align}\label{eq:mean-RxPi1}
	\mu_{p_r}&= p_d\left[\mathbb{E}_{\widehat{\mathbf{h}}_{\mathrm{A}}} \left\lbrace\frac{\widehat{\mathbf{h}}_{\mathrm{A}}^{\mathrm{H}}\,\mathbf{C}_{\mathbf{h}\big|\widehat{\mathbf{h}}_{\mathrm{A}}}\widehat{\mathbf{h}}_{\mathrm{A}}}{\norm{\widehat{\mathbf{h}}_{\mathrm{A}}}^2}+\frac{\widehat{\mathbf{h}}_{\mathrm{A}}^{\mathrm{H}}\,\boldsymbol{\mu}_{\mathbf{h}\big|\widehat{\mathbf{h}}_{\mathrm{A}}} \boldsymbol{\mu}_{\mathbf{h}\big|\widehat{\mathbf{h}}_{\mathrm{A}}}^{\mathrm{H}} \widehat{\mathbf{h}}_{\mathrm{A}}}{\norm{\widehat{\mathbf{h}}_{\mathrm{A}}}^2}\right\rbrace\right]\nonumber\\
	&\approx p_d \left[\mathbb{E}_{\widehat{\mathbf{h}}_{\mathrm{A}}}  \left\lbrace\mathbf{C}_{\mathbf{h}\big|\widehat{\mathbf{h}}_{\mathrm{A}}} \right\rbrace+
	\mathbb{E}_{\widehat{\mathbf{h}}_{\mathrm{A}}} \left\lbrace\boldsymbol{\mu}_{\mathbf{h}\big|\widehat{\mathbf{h}}_{\mathrm{A}}} \boldsymbol{\mu}_{\mathbf{h}\big|\widehat{\mathbf{h}}_{\mathrm{A}}}^{\mathrm{H}}\right\rbrace\right] 
	\nonumber\\
	&= p_d \left[\norm{\boldsymbol{\mu}_{\mathbf{h}}}^2+\frac{\beta\,\sigma_{\mathrm w}^{2}\left(K+1\right)+\beta^2\,\mathrm{E}_c\,N}{\left(K+1\right)\left(\beta\,\mathrm{E}_c+\sigma_{\mathrm w}^{2}\left(K+1\right)\right)}\right].
	\end{align}
	Lastly, on denoting the approximation defined in \eqref{eq:mean-RxPi1} by  $\widehat{\mu}_{p_r}$, and making some simple rearrangements to it, the desired result for $\widehat{\mu}_{p_r}$, as given by \eqref{eq:mean-aprRxP},
	can be obtained. \textcolor{black}{Here, we would like to highlight that since $\mathbf{C}_{\mathbf{h}\big|\widehat{\mathbf{h}}_{\mathrm{A}}}$ and $\mathbb{E}_{\widehat{\mathbf{h}}_{\mathrm{A}}} \left\lbrace\boldsymbol{\mu}_{\mathbf{h}\big|\widehat{\mathbf{h}}_{\mathrm{A}}} \boldsymbol{\mu}_{\mathbf{h}\big|\widehat{\mathbf{h}}_{\mathrm{A}}}^{\mathrm{H}}\right\rbrace$ are not dependent on the API-influenced terms $\left(\mathbf{F}_{\mathrm A} ,\mathbf{C}_{\widehat{\mathbf{h}}_{\mathrm{A}}}\right)$, the approximated mean received power $\widehat{\mu}_{p_r}$  is also independent of them.}
\end{IEEEproof}
This tight approximation for mean received  power $\mu_{p_r}$ provides the following key insights. 
\begin{remark}\itshape
	Under the high-SNR regime with $\sigma_{\mathrm w}^{2}\to0$ and/or single antenna $\mathcal{S}$ scenario with $N=1$, the last term in \eqref{eq:mean-aprRxP} vanishes and the resulting $\widehat{\mu}_{p_r}$ approaches the maximum value of average received RF power $
	\mu_{p_r,\mathrm{id}}$ as defined in \eqref{eq:ideal} under the perfect CSI availability assumption. On the contrary, for low-SNR regime with $\beta N p_c\tau_c\ll\sigma_{\mathrm w}^{2}\left(K+1\right)$, the third term $\frac{\beta\,\sigma_{\mathrm w}^{2}\left(N-1\right)}{\beta N p_c\tau_c+\sigma_{\mathrm w}^{2}\left(K+1\right)}$ reduces to $\frac{\beta\left(N-1\right)}{K+1}$, and thereby the underlying $\widehat{\mu}_{p_r}$ approaches the mean received RF power $
	\mu_{p_r,\mathrm{iso}}$ as defined in \eqref{eq:iso}  for the isotropic transmission.
\end{remark}

\begin{corollary} \itshape
	For Rayleigh fading case with $K=0$ and $\boldsymbol{\mu}_{\mathbf{h}}=\mathbf{0}_N$, the tight approximation $\widehat{\mu}_{p_r}^{\mathrm{ray}}$ for the mean received RF power $\mu_{p_r}$ as obtained from \eqref{eq:mean-aprRxP} is given by
	\begin{align}\label{eq:mean-aprRxP-Ray}
	\widehat{\mu}_{p_r}^{\mathrm{ray}}\triangleq\,p_d \left[ \beta\,N -\frac{\beta\,\sigma_{\mathrm w}^{2}\left(N-1\right)}{\beta N p_c\tau_c+\sigma_{\mathrm w}^{2}}\right].
	\end{align} 
\end{corollary} 
\begin{remark}\itshape
	For the high-SNR regime ($\sigma_{\mathrm w}^{2}\to0$) and $N=1$ scenario,  $\widehat{\mu}_{p_r}^{\mathrm{ray}}$ approaches the mean received RF power $\mu_{p_r,\mathrm{id}}^{\mathrm{ray}}$ under perfect CSI availability. However, for the low-SNR regime with $\beta N p_c\tau_c\ll\sigma_{\mathrm w}^{2}$, $\widehat{\mu}_{p_r}^{\mathrm{ray}}$ reduces to the mean received RF power $\mu_{p_r,\mathrm{iso}}^{\mathrm{ray}}$ for the isotropic transmission.
\end{remark}

\section{Joint Optimal Power and Time Allocation}\label{sec:Joint}
To maximize the efficiency of the proposed DL hybrid EBF using the UL LS-based CE, we need to maximize the stored energy $\mathrm{E}_s\triangleq \mathrm{E}_h-\mathrm{E}_c$ at $\mathcal{U}$ in each block. Since the cost of UL CE is in terms of energy consumption during the pilot signal transmission from $\mathcal{U}$, the average stored energy $\mu_{\mathrm{E}_h}$ at $\mathcal{U}$, as available after replenishing the consumed energy in CE, is given by
\begin{align}\label{eq:apr-mu}
\mu_{\mathrm{E}_s}\triangleq&\,\mathbb{E}\left\lbrace \mathrm{E}_s\right\rbrace=\left(\tau-N\tau_c\right)\mu_{p_h}-N\tau_c\,p_c 
\nonumber\\
\approx&\,\widehat\mu_{\mathrm{E}_s}\triangleq\left(\tau-N\tau_c\right)\left[\mathcal{A}_{\widehat{i_0}}\,\widehat{\mu}_{p_r}+\mathcal{B}_{\widehat{i_0}}\right]-N\tau_c\,p_c,
\end{align}
where $\widehat{i_0}\triangleq\left\lbrace  i\mathrel{}\middle|\mathrel{}\widehat{\mu}_{p_r}\in\left[p_{\mathrm{th}_i},p_{\mathrm{th}_{i+1}}\right],\,1\le i\le L \right\rbrace$ and $\widehat{\mu}_{p_r}$, defined in \eqref{eq:mean-aprRxP}, is a function of $p_c$ and $\tau_c$. \textcolor{black}{Here, recalling from Lemma~\ref{lem:RV-RxRFP}, under the practically motivated approximation for $\widehat{\mu}_{p_r}$, the average stored energy to be maximized is independent of the unknown API-dependent parameters.}
Next, we formulate the problem for jointly optimizing the PA $p_c$ at $\mathcal{U}$ for UL CE and TA $\tau_c$ for CE phase to maximize this stored energy $\widehat\mu_{\mathrm{E}_s}$. Then we obtain both joint and individually global optimal solutions after proving the generalized convexity~\cite{generalized} of the underlying problems.

\subsection{Optimization Formulation}\label{sec:formulation}
The above-mentioned desired goal can be mathematically formulated as below
\begin{equation*}\label{eqOPJ}
\begin{split}
\mathcal{OP}:&\;  \underset{p_c,\,\tau_c}{\textrm{maximize}} \;  \widehat\mu_{\mathrm{E}_s},\qquad\textrm{subject to}\;\;({\rm C1}):p_c\le p_{\max},\nonumber\\
& ({\rm C2}):p_c\ge 0,\quad({\rm C3}):N\tau_c\le \tau,\quad ({\rm C4}):\tau_c\ge 0.
\end{split}
\end{equation*}
In general $\mathcal{OP}$ is nonconvex because though the constraints are linear, the objective $\widehat\mu_{\mathrm{E}_s}$ involves the coupled term $\mathrm{E}_c$ containing the product of $p_c$ and $\tau_c$. However, in the following sections we show that after exploiting the convexity of the individual PA and TA optimization, the jointly global optimal solution $\left(p_c^*,\,\tau_c^*\right)$ for $\mathcal{OP}$ can be derived in closed form. 

\textcolor{black}{Before we proceed further, it is worth noting that all the optimization related computations are carried out at $\mathcal{S}$ using the LSE obtained using the proposed hybrid CE protocol and only the statistical (expectation) knowledge of $\mathbf{h}$ is needed.  In fact, we don't need exact phasor information for the spectral components and instead only require to know  $\norm{\boldsymbol{\mu}_{\mathbf{h}}}^2$ at $\mathcal{S}$, which on ignoring API compensation and considering zero mean  AWGN can be easily obtained by taking the average over  received pilot signals. Further, this statistic remains good for several coherence blocks for static node deployment scenarios like ours and it is a common assumption used in similar investigations on optimizing RFET efficiency over Rician fading channels~\cite{CSI-SU-WET,ICASSP18,Signal-WPT-TC17}.}

\subsection{Optimal Power Allocation at EH User for given TA $\tau_c$}
For a given TA $\tau_c$ for each CE sub-phase, $\mathcal{OP}$ reduces to
\begin{equation*}\label{eqOPP} 
\mathcal{OP}1:\quad\underset{p_c}{\textrm{maximize}} \;\; \widehat\mu_{\mathrm{E}_s},\qquad\textrm{subject to}\quad({\rm C1}),\;({\rm C2}). 
\end{equation*}
$\mathcal{OP}1$ is a convex problem because for a given $\tau_c$, $\widehat\mu_{\mathrm{E}_s}$ is strictly concave in $p_c,\,\forall i=1,2,\ldots, L,$ as shown below
\begin{eqnarray}\label{eq:conc-p}
\frac{\partial^2\widehat\mu_{\mathrm{E}_s}}{\partial p_c^2}=-\frac{2\mathcal{A}_i\,p_d\left(\tau-N\tau_c\right)\sigma_{\mathrm w}^{2}N^2\left(N-1\right)\tau_c^2\beta^3}{\left(\beta N p_c\tau_c+\sigma_{\mathrm w}^{2}\left(K+1\right)\right)^3}<0.
\end{eqnarray}
Using this, the global optimal solution of $\mathcal{OP}1$ is defined next.
\begin{lemma}\label{lem:OPA} 
	\textit{The global optimal PA $p_c$ for a given TA $\tau_c$, is}
	\begin{subequations}
		\begin{equation}\label{eq:opt-pc}
		p_c^*\triangleq\min\{\max\{0,\,p_{c_{i_p}}\},\,p_{\max}\},\quad\textit{where},
		\end{equation} 
		\begin{eqnarray}\label{eq:pOi}
		i_p\triangleq\min\left\lbrace i\mathrel{}\middle|\mathrel{}\widehat{\mu}_{p_r}\Big|_{p_c=p_{c_i}}\in\left[p_{\mathrm{th}_i},p_{\mathrm{th}_{i+1}}\right],1\le i\le L\right\rbrace,
		\end{eqnarray} 
		\begin{eqnarray}\label{eq:opt-pci}
		p_{c_i}\triangleq\frac{ \sqrt{\mathcal{A}_i\,p_d\left(\tau-N\tau_c\right)\sigma_{\mathrm w}^{2}\left(N-1\right)} -\frac{\sigma_{\mathrm w}^{2}\left({K+1}\right)}{
				\beta}}{ N \tau_c}.
		\end{eqnarray} 
	\end{subequations}
\end{lemma} 
\begin{IEEEproof}
	Using \eqref{eq:conc-p}, the global optimal $p_c$ for a given TA $\tau_c$ can be obtained from \eqref{eq:apr-mu} on solving $\frac{\partial\widehat\mu_{\mathrm{E}_s}}{\partial p_c}=0$ in $p_c$. As $\widehat\mu_{\mathrm{E}_s}$ in \eqref{eq:conc-p} takes $L$ different nonzero values, $\frac{\partial\widehat\mu_{\mathrm{E}_s}}{\partial p_c}=0$ takes $L$ solutions in $p_c$ as denoted by $p_{c_i}$ defined in \eqref{eq:opt-pci}. However, among these $L$ potential candidates $p_{c_i}$ for $p_c^*$, there is only one feasible linear piece with index $i=i_p$ that uniquely represents the global maximum value of $\widehat\mu_{\mathrm{E}_s}$ in $p_c$ for a given $\tau_c$. Hence, the optimal index, as denoted by $i_p$, can be defined as the first linear piece $i$ among the $L$ pieces such that the corresponding  $\widehat{\mu}_{p_r}$ as defined in \eqref{eq:mean-aprRxP} with  $p_c=p_{c_i}$ lies between $p_{\mathrm{th}_{i}}$ and $p_{\mathrm{th}_{i+1}}$. The fact that only one PA $p_{c_i}$ satisfies the underlying $p_{\mathrm{th}_i}\le\widehat{\mu}_{p_r}\le p_{\mathrm{th}_{i+1}}$  requirement can be observed from the strict concavity of $\widehat\mu_{\mathrm{E}_s}$ in $p_c$. Hence, as an increasing index $i\in\left[1,L\right]$ implies higher received RF power $\widehat{\mu}_{p_r}$, objective $\widehat\mu_{\mathrm{E}_s}$ in $\mathcal{OP}1$ is either strictly increasing for initial pieces having indices $1\le i< i_p$, then strictly concave for the $i_p$th piece, and then strictly decreasing for the indices $i_p< i\le L$. This index $i_p$ uniquely characterizing the optimal linear piece $i$ defining the global optimal $p_c^*$ along with the bounds on $p_c\in\left[0,\,p_{\max}\right]$, completes the proof. 
\end{IEEEproof}

\subsection{Optimal Time Allocation for a given PA $p_c$}
The mathematical formulation for the obtaining optimal TA $\tau_c$ for  PA $p_c$ at $\mathcal{U}$ is define below:
\begin{equation*}\label{eqOPT}
\begin{split}
\mathcal{OP}2:&\;\; \underset{\tau_c}{\textrm{maximize}} \;\; \widehat\mu_{\mathrm{E}_s},\quad\textrm{subject to}\quad({\rm C3}),\;({\rm C4}).
\end{split}
\end{equation*}
Here, $\mathcal{OP}2$ is convex because it involves a strictly-concave objective $\widehat\mu_{\mathrm{E}_s}$, satisfying $\frac{\partial^2\widehat\mu_{\mathrm{E}_s}}{\partial \tau_c^2}<0,\forall i\le L,$ as shown below, and all the constraints are linear functions of $\tau_c$,
\begin{align}\label{eq:conc-TA}
\frac{\partial^2\widehat\mu_{\mathrm{E}_s}}{\partial \tau_c^2}=\frac{-2\mathcal{A}_i\,p_d\,p_c\sigma_{\mathrm w}^{2}\beta^2 \left(\beta  p_c \tau+\sigma_{\mathrm w}^{2}\left(K+1\right)\right)}{\left[N^2\left(N-1\right)\right]^{-1}\left(\beta N p_c\tau_c+\sigma_{\mathrm w}^{2}\left(K+1\right)\right)^3}<0.
\end{align}
Therefore, the optimal TA $\tau_{c_i}$, for a given PA $p_c$ and  PWLA parameters $\mathcal{A}_i,\mathcal{B}_i,\,\forall i=1,2,\ldots, L$, is obtained as:
\begin{eqnarray}\label{eq:opt-tau}
\tau_{c_i}\triangleq\frac{\sqrt{\frac{ \left(p_c \tau+\frac{\sigma_{\mathrm w}^{2}}{\beta}\left({K+1}\right)\right)\sigma_{\mathrm w}^{2}\left(N-1\right)}{ \frac{\beta\,N}{K+1}+\norm{\boldsymbol{\mu}_{\mathbf{h}}}^2 +\frac{\mathcal{B}_i+p_c}{\mathcal{A}_i p_d}}}-\frac{\sigma_{\mathrm w}^{2}}{\frac{\beta}{K+1}}}{N p_c}.
\end{eqnarray}
Using this result,  the optimal solution of $\mathcal{OP}2$ can be characterized via Corollary~\ref{cor:OTA}.
\begin{corollary}\label{cor:OTA}
	\textit{The global optimal TA for PA $p_c$, is given by}
	\begin{align}\label{eq:opt-tc}
		&\tau_c^*\triangleq\min\{\max\{0,\,\tau_{c_{i_\tau}}\},\,\frac{\tau}{N}\},\quad\textit{with }\nonumber\\ 
		&i_\tau \triangleq\min \left\lbrace i\mathrel{}\middle|\mathrel{}\widehat{\mu}_{p_r}\Big|_{\tau_c=\tau_{c_i}}\in\left[p_{\mathrm{th}_i},p_{\mathrm{th}_{i+1}}\right],i\in\left[1, L\right]\right\rbrace. 
	\end{align}  
\end{corollary} 
\begin{IEEEproof}
	Following the proof of Lemma~\ref{lem:OPA} and using the strict concavity of $\widehat\mu_{\mathrm{E}_s}$ in $\tau_c$ for a given $p_c$, the optimal $\tau_c$, as denoted by $\tau_{c_i}$, is defined in \eqref{eq:opt-tau}. Similar to uniqueness claim for $i_p$ as proved in Lemma~\ref{lem:OPA}, the  optimal index $i=i_\tau$ defines the only feasible $\tau_{c_i}$ satisfying the underlying $p_{\mathrm{th}_i}\le\widehat{\mu}_{p_r}\le p_{\mathrm{th}_{i+1}}$, and hence yielding the global optimal $\tau_{c_{i_\tau}}$. This along with the feasible $\tau_c^*$ to satisfy the boundary constraints $({\rm C3})$ and $({\rm C4})$ yields the  optimal TA in \eqref{eq:opt-tc}.
\end{IEEEproof}



\subsection{Proposed Global Optimization Algorithm}
Now we focus on jointly optimizing $p_c$ and $\tau_c$ in $\mathcal{OP}$. In contrast to $\mathcal{OP}1$  and $\mathcal{OP}2$, $\mathcal{OP}$ is nonconvex. However, via Theorem~\ref{th:GOS} we show that exploiting the collective impact of $p_c$ and $\tau_c$ in terms of energy consumption $\mathrm{E}_c$ during CE yields the  jointly global optimal solution for $\mathcal{OP}$. 

\begin{theorem}\label{th:GOS}\itshape
	The global optimal solution $\left(p_{c_J}^*,\tau_{c_J}^*\right)$ for $\mathcal{OP}$, yielding maximum $\widehat\mu_{\mathrm{E}_s}$ at $\mathcal{U}$, is
	\begin{subequations} 
		\begin{equation}\label{eq:jOPA}
		p_{c_J}^*=p_{\max},\qquad\,\tau_{c_J}^*=\tau_{c_{i^*}}\; \text{ with } p_c=p_{\max} \text{ in } \eqref{eq:opt-tau},
		\end{equation}
		\begin{align}\label{eq:jOTA}
		i^*\triangleq\min\left\lbrace i\mathrel{}\middle|\mathrel{}\widehat{\mu}_{p_r}\Big|_{\underset{\tau_c=\tau_{c_i},}{p_c=p_{\max}}}\in\left[p_{\mathrm{th}_i},p_{\mathrm{th}_{i+1}}\right],i\in\left[1,L\right]\right\rbrace.
		\end{align}
	\end{subequations}
\end{theorem} 
\begin{IEEEproof}
	As the joint optimal solution is defined by $p_{c_J}^*,\tau_{c_J}^*,$ and $ 
	i^*$, as given in \eqref{eq:jOPA} and \eqref{eq:jOTA}, we present the proof for these three expressions in the next three separate paragraphs.
	
	From \eqref{eq:Es} and \eqref{eq:apr-mu}, $\widehat\mu_{\mathrm{E}_s}=\left(\tau-N\tau_c\right)\left[\mathcal{A}_{\widehat{i_0}}\,\widehat{\mu}_{p_r}+\mathcal{B}_{\widehat{i_0}}\right]-\mathrm{E}_c=\widehat\mu_{\mathrm{E}_h}-\mathrm{E}_c$. With $\mathrm{E}_c=N p_c \tau_c$, the average received power  $\widehat{\mu}_{p_r}$, as defined in \eqref{eq:mean-aprRxP}, is strictly increasing in both $p_c$ and $\tau_c$. Therefore, as both $p_c$ and $\tau_c$ have an identical effect on $\mathrm{E}_c$, and $\widehat\mu_{\mathrm{E}_s}$  is strictly increasing in $\widehat{\mu}_{p_r}$, the optimal PA $p_c$ for the CE phase should be such that it maximizes the monotonically increasing $\widehat\mu_{\mathrm{E}_h}$ (cf.  \eqref{eq:Es}). Hence, the optimal 
	PA is $p_c^*=p_{\max}$. Also, between $p_c$ and $\tau_c$, $p_c$ was chosen to be set to its maximum value,  because unlike its variation in  $p_c$, $\widehat\mu_{\mathrm{E}_h}$ is not strictly increasing in $\tau_c$. 
	
	Now with $p_c$ set to $p_{\max}$,  $\tau_c$ can now be optimized to in turn optimize $\mathrm{E_c}$ that maximizes $\widehat\mu_{\mathrm{E}_s}$. As from \eqref{eq:conc-TA} we note that for a given $i$ with $p_c=p_{\max}$, $\widehat\mu_{\mathrm{E}_s}$ is strictly concave in $\tau_c$, the optimal TA  is given by $\tau_{c_i}$, which was defined by \eqref{eq:opt-tau}.
	
	Lastly, we show that there is only one linear piece index $i\in\left[1,L\right]$ which yields $\tau_{c_i}$ that uniquely defines the global maximum value of $\widehat\mu_{\mathrm{E}_s}$. Hence, the optimal $i$, as denoted by $i^*$, can be defined as the first linear piece among the $L$ pieces such that the corresponding  $\widehat{\mu}_{p_r}$ as defined in \eqref{eq:mean-aprRxP} with  $\tau_c=\tau_{c_{i^*}},p_c=p_{\max}$ lies between $p_{\mathrm{th}_{i^*}}$ and $p_{\mathrm{th}_{i^*+1}}$. The uniqueness of $i^*$ can be guaranteed from the fact that for each linear piece,  $\widehat\mu_{\mathrm{E}_s}$ is strictly concave in $\tau_c$  with $p_c=p_{\max}$, i.e., $\frac{\partial^2\widehat\mu_{\mathrm{E}_s}}{\partial \tau_c^2}<0, \forall i$. Proceeding similar to the proof of Lemma~\ref{lem:OPA}, $\widehat\mu_{\mathrm{E}_s}$ can be shown to be strictly increasing for initial linear pieces having indices $1\le i< i^*$, then strictly concave for the $i^*$th piece, and strictly decreasing for $i^*< i\le L$. This completes the proof. 
\end{IEEEproof}

\begin{algorithm}[!t]	
{\small		
\caption{\small Jointly global optimal  $p_c$ and $\tau_c$ maximizing $\widehat\mu_{\mathrm{E}_h}$}\label{Algo:Opt}		
\begin{algorithmic}[1]
	\setstretch{1.1}			
	\Require Goal (optimal PA, TA, or joint), system and channel  parameters, along with fixed PA $p_{c_0}$ and fixed TA $\tau_{c_0}$ 
	\Ensure Optimal PA $p_c^*$ and TA $\tau_c^*$ for CE phase along with $\widehat\mu_{\mathrm{E}_s}^*$  			
	\For{$i \in \{1,2,\dots,L\}$}			
	\If {$\mathcal{OP}1$ has to be solved for optimal PA}			
	\State Obtain $p_{c_i}$ using \eqref{eq:opt-pc} 
	\State Set $\widehat{\mu}_{p_{r_i}}=\widehat{\mu}_{p_r}$ using \eqref{eq:mean-aprRxP} with $p_c=p_{c_i}$ and $\tau_c=\tau_{c_0}$ 	
	\Else				
	\State Obtain $\tau_{c_i}$ using \eqref{eq:opt-tau} 
	\If {$\mathcal{OP}2$ has to be solved for optimal TA}		
	\State Set $\widehat{\mu}_{p_{r_i}}=\widehat{\mu}_{p_r}$ using \eqref{eq:mean-aprRxP} with $p_c=p_{c_0}$ and $\tau_c=\tau_{c_i}$	
	\Else
	\State Set $\widehat{\mu}_{p_{r_i}}=\widehat{\mu}_{p_r}$ using \eqref{eq:mean-aprRxP} with $p_c=p_{\max},\,\tau_c=\tau_{c_i}$			 
	\EndIf			
	\EndIf	
	
	\If {$p_{\mathrm{th}_{i}}\le\widehat{\mu}_{p_{r_i}}\le p_{\mathrm{th}_{i+1}}$}
		\If {$\mathcal{OP}1$ has to be solved for optimal PA}		
			\State Set $i_p=i$,\;\; $p_c^*=p_{c_{i_p}}$,\; and\; $\tau_c^*=\tau_{c_0}$ 
	    \ElsIf {$\mathcal{OP}2$ has to be solved for optimal TA}
		    \State Set $i_\tau=i$,\;\; $\tau_c^*=\tau_{c_{i_\tau}}$,\; and\; $p_c^*=p_{c_0}$	
		\Else	
			\State Set  $i^*=i$,\;\; $p_c^*=p_{\max}$,\; and\; $\tau_c^*=\tau_{c_{i^*}}$ 				
		\EndIf 		
		\State Obtain maximum $\widehat\mu_{\mathrm{E}_h}^*$ using \eqref{eq:apr-mu} with $p_c=p_c^*,\,\tau_c=\tau_c^*$
		\State \textbf{break}
	\EndIf 		
	\EndFor		 	
\end{algorithmic}		
}	
\end{algorithm}


 
The step-by-step procedure to obtain both individual (PA or TA) and joint optimization results is summarized in Algorithm~\ref{Algo:Opt}. It shows that the joint or individual global optimal PA and TA solution can be obtained in closed form by selecting the best among the $L$ possible candidates. This corroborates the low computational complexity of the proposed jointly global optimal design that incorporates the nonlinear RF EH model and takes into account practical  API and CE errors.

\section{Numerical Performance Evaluation}
\label{sec:results}
Here, we numerically evaluate the optimized performance of the proposed hybrid EBF protocol under API and CE errors. Unless otherwise stated explicitly, in the figures that follow we have set $N=20$, $\tau=10$ms, $\tau_{c_0}=\frac{\tau}{1000}=10\mu$s, $\Delta_{g_{i_k}}=\Delta_{\phi_{i_k}}=\Phi_{g_{i_k}}=\Phi_{\phi_{i_k}}=\Delta,\forall i\in\mathcal{N},\,k=1,2,$ $p_d=36$dBm, $p_{\max}=10$dBm, $p_{c_0}=\frac{p_{\max}}{100}=-10$dBm, $\sigma_{\mathrm w}^{2}=-150$dBm, $\Delta=0.065$, $\delta=\frac{3\times 10^8}{2f_c}, \psi=0^{\circ}, K=2,\alpha_i=1$ and $\beta=\frac{\varpi}{d^{\varrho}}$, where $\varpi=\left(\frac{\delta}{2\pi}\right)^2$ being the average channel attenuation at unit reference distance with $f_c=915$MHz \cite{RFEH2008} being $\mathcal{S}$ frequency, $d=15$m is $\mathcal{S}$-to-$\mathcal{U}$ distance, and $\varrho=2.5$ is the path loss exponent. The values for fixed TA $\tau_{c_0}$ and PA $p_{c_0}$ have been selected so as to ensure that $\mu_{\mathrm{E}_s}$ in \eqref{eq:apr-mu} is positive. For incorporating the practical nonlinear RF-to-DC conversion operation at $\mathcal{U}$, the rectification efficiency function $\mathcal{L}\left\lbrace\cdot\right\rbrace$ is modeled using \eqref{eq:PWLA0} with parameters $\mathcal{A}_i,\mathcal{B}_i,p_{\mathrm{th}_i},\forall i\in\left[1,L\right],L=5,$ as defined in Section~\ref{sec:RFEH} for RF EH circuit designed in~\cite{RFEH2008} for efficient far-field (i.e., long range) RFET . Lastly, all the simulation results plotted here have been obtained after taking average over $10^5$ independent channel realizations.

\subsection{Validation of Analysis}\label{sec:valid_ana} 
\begin{figure}[!t] 
	\centering\includegraphics[width=3.45in]{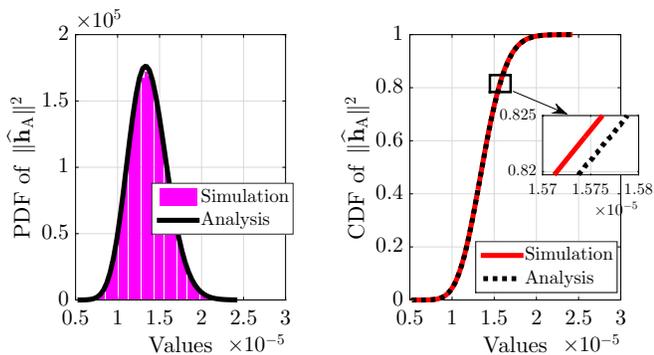}
	\caption{Validating distribution  of the squared-norm $\norm{\widehat{\mathbf{h}}_{\mathrm{A}}}^2$ of proposed LSE.}
	\label{fig:PDF-CDF}  
\end{figure}

Here, we first validate the CE analysis carried out in Section~\ref{sec:CE}. In particular, the distribution of $\norm{\widehat{\mathbf{h}}_{\mathrm{A}}}^2$ as derived in Section~\ref{sec:normLSE}, with statistics defined by \eqref{eq:varS-LSE} and \eqref{eq:3-apprC}, using practical API based approximation~\eqref{eq:approx1} is verified via extensive Monte Carlo simulations in Fig.~\ref{fig:PDF-CDF}. This is a key metric which is used for deriving the average stored energy $\widehat\mu_{\mathrm{E}_h}$ at $\mathcal{U}$ after replenishing the energy expenditure $\mathrm{E}_c$ during the CE phase. Both analytical PDF and CDF of $\norm{\widehat{\mathbf{h}}_{\mathrm{A}}}^2$ have been validated against their corresponding \textit{simulated} values. From Fig.~\ref{fig:PDF-CDF}, a very close match between the analysis and simulation can be observed. Further, the RMSE value $0.0016$ (very close to 0) and R-square statistics  value $0.9999$ (very close to 1) signify the goodness~\cite{hooper2008structural} of the proposed  analytical approximation for quantifying the distribution of $\norm{\widehat{\mathbf{h}}_{\mathrm{A}}}^2$.

\quad\;
\begin{figure}[!t] 
	\centering\includegraphics[width=3.45in]{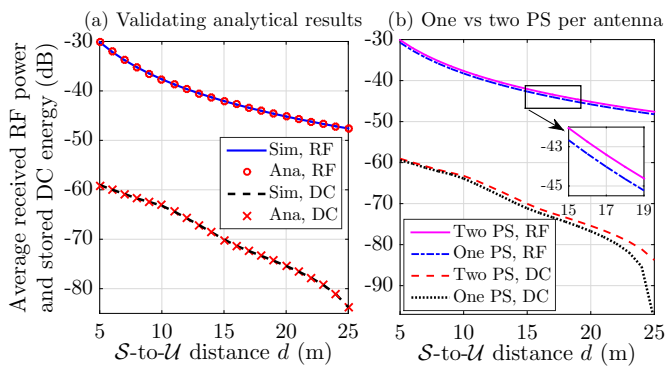}
	\caption{\textcolor{black}{Verifying analytical approximations for  $\widehat{\mu}_{p_r}$ and $\widehat\mu_{\mathrm{E}_s}$, along with the significance of having two DCPS per antenna.}}
	\label{fig:RF-HD}   
\end{figure}  
 
Now, we validate the quality of other two approximations used for obtaining the closed-form expression for the average stored energy at $\mathcal{U}$. The first approximation is based on the practical limits on the value of API-based parameter $\Delta$  as outlined in Section~\ref{sec:aprACE}. To validate this,  we have compared the variation of the average received RF power $\mu_{p_r}$ at $\mathcal{U}$, as obtained by simulating \eqref{eq:mean-RxP}, with increasing $d$ in  \textcolor{black}{Fig.~\ref{fig:RF-HD}(a)} against its  tight analytical approximation $\widehat{\mu}_{p_r}$ as defined by \eqref{eq:mean-aprRxP}. The underlying RMSE of $0.0039$ between $\mu_{p_r}$ and $\widehat{\mu}_{p_r}$ validates the quality~\cite{hooper2008structural} of this approximation~\eqref{eq:approx1}. \textcolor{black}{With this validation, we next investigate  the quality of the Jensen inequality based second approximation $\widehat\mu_{\mathrm{E}_s}$, as defined in \eqref{eq:apr-mu}, for the average stored DC energy $\mu_{\mathrm{E}_s}=\left(\tau-N\tau_c\right)\mu_{p_h}-N\tau_c\,p_c$ obtained after simulating \eqref{eq:Eh-exact}. The tight match between  $\widehat\mu_{\mathrm{E}_s}$ and $\mu_{\mathrm{E}_s}$ as observed in \textcolor{black}{Fig.~\ref{fig:RF-HD}(a)} is also corroborated with the underlying RMSE $<0.082$. Here, notice that this gap between the average stored energies $\widehat\mu_{\mathrm{E}_s}$ and $\mu_{\mathrm{E}_s}$ is higher than the corresponding average RF powers $\mu_{p_r}$ and $\widehat{\mu}_{p_r}$ because the former involves errors due to both approximations. However, the quality of these  approximations is acceptable for practical settings investigated in the following sections. Furthermore, as the average stored energy $\mu_{\mathrm{E}_s}$ defined in \eqref{eq:apr-mu} is a positive linear transformation of the mean harvested DC power $\mu_{ p_h }$ defined in \eqref{eq:PWLA1}, by providing validation for $\mu_{\mathrm{E}_s}$ via \textcolor{black}{Fig.~\ref{fig:RF-HD}(a)} we also in turn   validate analytical result $\widehat{\mu}_{ p_h}$ derived for $\mu_{ p_h }$.}

\textcolor{black}{Here, we also verify the practical significance of having two PS (or, in particular, two DCPS with a combiner~\cite{HBF-Prac1}) for each antenna element at $\mathcal{S}$ to achieve the full digital EBF gains via the adopted hybrid architecture. Specifically, we plot the variation of average received RF power  and effective stored  energy at $\mathcal{U}$ due to hybrid EBF with both one and two DCPS per antenna element at $\mathcal{S}$ in Fig.~\ref{fig:RF-HD}(b). Here, to maximize the EBF gains, the analog precoder for single PS case is set as $\mathrm{exp}\left\lbrace{-j\,\phase{\mathbf{z}_{\mathrm{A}}^*}}\right\rbrace$~\cite{ICASSP19_PIS}. On averaging the performance over different RFET ranges $d$, we  notice that the received RF power   and effective stored  energy  respectively  get reduced by  $14$\% and $23$\% when only one DCPS is employed for each antenna element at $\mathcal{S}$. This performance loss becomes concerningly  higher for larger arrays ($N\gg1$) and longer ranges ($d>20$m).}
 
\begin{figure}[!t] 	
	\centering\includegraphics[width=3.45in]{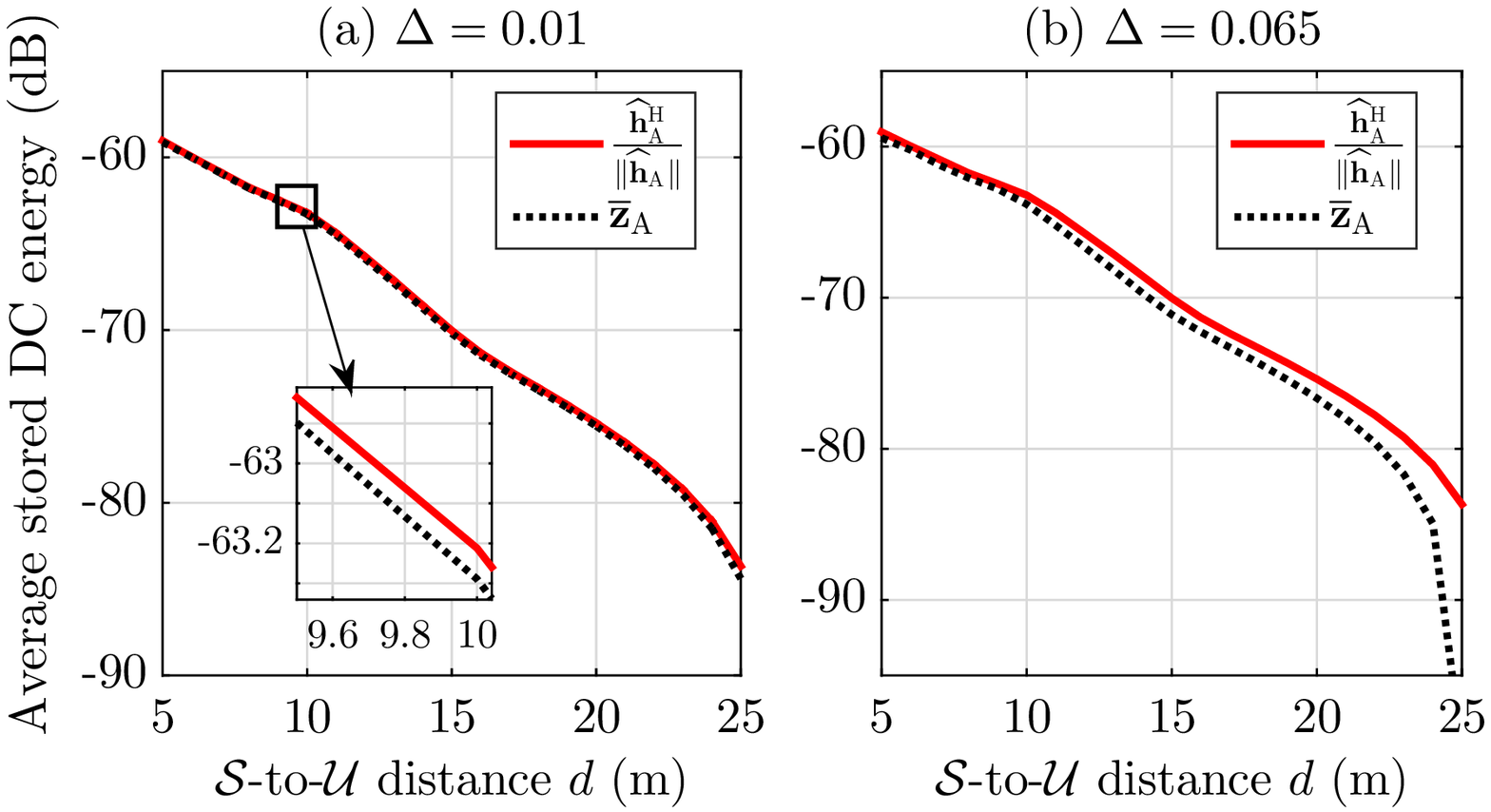}
	\caption{Comparing average stored energy under                                                                                                                                                                                                                                                                                                                                                                                                                                                                                                                                                                                                                                                                                                                                                                                                                                                                                                                                                                                                                                                                                                                                                                                                                                                                                                                                                                                                                                                                                                                                                                                                                                                                                                                                                                                                                                                                                                                                                                                                                                                                                                                                                                                                                                                                                                                                                                                                                                                                                                                                                                                                                                                                                                                                                                                                                                                                                                                                                                                                                                                                                                                                                                                                                                                                                                                                                                                                                                                                                                                                                                                                                                                                                                                                                                                                                                                                                                                                                                                                                                                                                                                                                                                                                                                                                                                                                                                                                                                                                                                                                                                                                                                                                                                                                                                                                                                                                                                                                                                                                                                                                                                                                                                                                                                                                                                                                                                                                                                                                                                                                                                                                                                                                                                                                                                                                                                                                                                                                                                                                                                                                                                                                                                                                                                                                                                                                                                                                                                                                                                    two precoding designs: $\mathbf{z}_{\mathrm{A}}=\frac{\widehat{\mathbf{h}}_{\mathrm{A}}^{\mathrm{H}}}{\norm{\widehat{\mathbf{h}}_{\mathrm{A}}}}$  and $\overline{\mathbf{z}}_{\mathrm{A}}=\Theta\left\lbrace\mathbf{z}_{\mathrm{A}}\right\rbrace$ as defined by \eqref{eq:preC}.}
	\label{fig:HDC-A-P} 
\end{figure} 
Lastly, in Fig.~\ref{fig:HDC-A-P} we investigate the performance degradation due to the approximation of using $\mathbf{z}_{\mathrm{A}}$ instead of  $\overline{\mathbf{z}}_{\mathrm{A}}$ as mentioned in  Sections~\ref{sec:RxPRV-appr} and~\ref{sec:TxEBF}. We note that for $\Delta=0.01$, the average stored energy $\widehat\mu_{\mathrm{E}_s}$ performance closely follows the corresponding simulated values for average stored energy with analog precoder $\overline{\mathbf{z}}_{\mathrm{A}}$ influenced by API. However, for high communication ranges $d>10$m with $\Delta=0.065$, the performance degradation due to API in analog precoder design can be clearly observed. Since, unlike CE errors, the performance degradation due to API cannot be compensated with increasing energy allocation $\mathrm{E}_c$ for the CE phase, and novel API estimation and compensation protocols are needed which are out of scope of this work, this gap cannot be eliminated. Thus, under this practical limitation, we focus on the joint resource allocation for maximizing the stored energy under CE errors in the API-affected LSE for $\mathbf{h}_{\mathrm{A}}$.

\begin{figure}[!t] 
	\centering\includegraphics[width=3.45in]{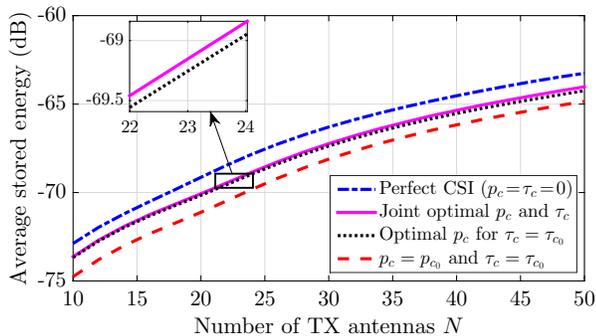}
	\caption{The average stored energy  $\widehat\mu_{\mathrm{E}_s}$ performance at $\mathcal{U}$ with increasing number of antennas $N$ at $\mathcal{S}$.}
	\label{fig:N} 
\end{figure} 

\subsection{Impact of Key System Parameters}\label{sec:insights}
Now we investigate the impact of four key system parameters, namely, (a) number of antennas $N$ at $\mathcal{S}$, (b) Rice factor $K$, (c) communication range $d$, and (d) API severity parameter $\Delta$. For each case, the average stored energy at $\mathcal{U}$ for the `ideal' scenario with perfect CSI availability is compared against the three practical scenarios, suffering from API and CE errors, which include: (i) joint optimal PA-TA, (ii) optimal PA $p_c^*$ for fixed TA $\tau_c=\tau_{c_0}$, and (iii) fixed $p_c=p_{c_0},\,\tau_c=\tau_{c_0}$.

\begin{figure}[!t]		
	\centering\includegraphics[width=3.45in]{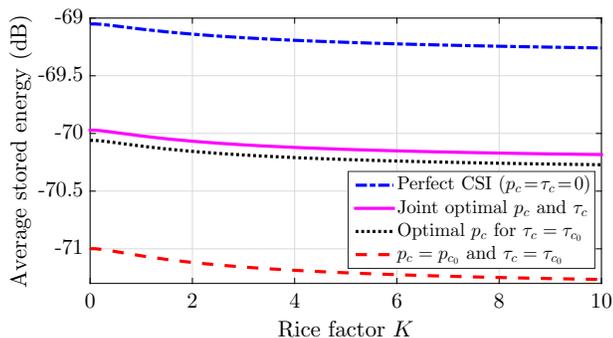} 
	\caption{Impact of increasing Rice factor $K$ values on   $\widehat\mu_{\mathrm{E}_s}$.}
	\label{fig:K} 
\end{figure} 
First, while investigating the average stored energy $\widehat\mu_{\mathrm{E}_s}$ variation with $N$ in  Fig.~\ref{fig:N}, we  observe that the optimal PA can yield noticeable performance improvement over the fixed scheme. However, the further enhancement obtained by jointly optimizing PA and TA is not significant over as achieved optimizing  PA alone. Hence, we note that the proposed joint optimal PA and TA can help in enhancing the practical hybrid EBF performance so that it can reach closer to the theoretical limit as achieved by the perfect CSI availability case that does not require any PA or TA for CE, i.e., has $p_c=\tau_c=0$. Another key observation from Fig.~\ref{fig:N} is that the performance enhancement in $\widehat\mu_{\mathrm{E}_s}$ as achieved by joint optimization over optimal PA alone gets slightly increased with higher number of antenna elements $N$ at $\mathcal{S}$.

Second, we investigate the underlying variation with Rice factor $K$ in Fig.~\ref{fig:K}. Though this variation with $K$ is not very significant, the average stored energy $\widehat\mu_{\mathrm{E}_s}$ decreases with increasing $K$.  However, with $K$ increasing from $0$ to $10$, the underlying $\widehat\mu_{\mathrm{E}_s}$ only decreases by $0.25$ dB.  Again, here also optimal PA, clearly outperforming the fixed allocation scheme, closely follows the performance of joint optimization scheme.

\begin{figure}[!t]	
	\centering\includegraphics[width=3.45in]{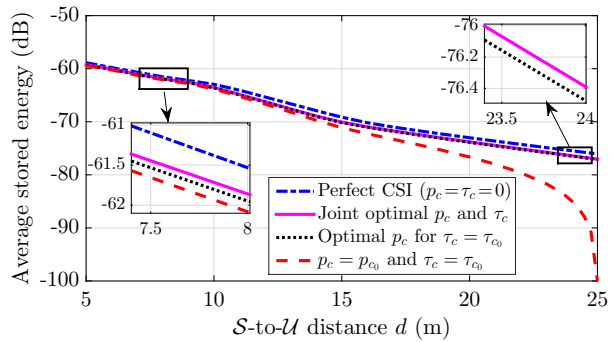} 
	\caption{Degradation in $\widehat\mu_{\mathrm{E}_s}$ with increasing $\mathcal{S}$-to-$\mathcal{U}$  range $d$.}
	\label{fig:d} 
\end{figure}

Next, we plot the variation of $\widehat\mu_{\mathrm{E}_s}$ in $\mathcal{S}$-to-$\mathcal{U}$ distance $d$ in Fig.~\ref{fig:d}. This variation of $d$ has been bounded  by the maximum communication range $d=25$m satisfying the received energy sensitivity constraint  of RF EH circuit~\cite{RFEH2008}. This requirement implies that the input RF power at $\mathcal{U}$ should be more than $-22$dBm for having nonzero harvested DC power to be nonzero after RF-to-DC rectification (cf. Fig.~\ref{fig:RFEH}). Here again, the same relative stored energy performance observed for the four cases investigated. However, we notice that for the larger values of $d$ the importance of optimization PA and TA becomes more critical because the stored energy $\widehat\mu_{\mathrm{E}_s}$  decreases very sharply over longer communication ranges $d$. Hence, though the performance gap between perfect CSI and the joint PA-TA increases from less than $0.5$dB to slightly more than $1$ dB when $d$ increases from $5$m to $25$m, this underlying gap between the joint PA-TA and fixed scheme increases from $0.2$dB to about a significant gain of $23$ dB.

\begin{figure}[!t] 	
	\centering\includegraphics[width=3.45in]{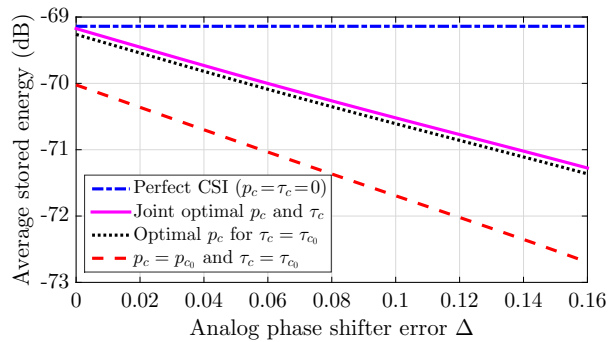}
	\caption{Variation of the average stored energy $\widehat\mu_{\mathrm{E}_s}$ with increasing values for the parameter $\Delta$ denoting API  severity.}
	\label{fig:API} 
\end{figure} 

\begin{figure}[!t] 
	\centering\includegraphics[width=3.45in]{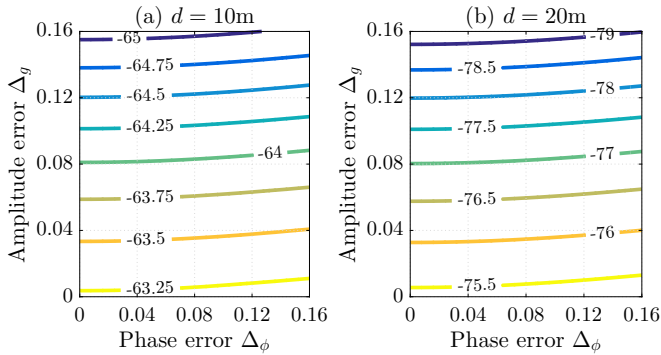}
	\caption{\textcolor{black}{Average stored energy  $\widehat\mu_{\mathrm{E}_s}$ performance in dB against the increasing $\Delta$ values for amplitude and phase errors.}}
	\label{fig:Amp_Phi}  
\end{figure}

Now, we investigate the fourth key API severity parameter $\Delta$ in Fig.~\ref{fig:API}, where $\Delta=0$  implies no API. Intuitively, the perfect CSI  based stored energy remains uninfluenced with $\Delta$ variation. However, for the other three practical cases suffering from the joint API and CE errors, the stored energy performance degrades with increasing $\Delta$ due to  the higher severity of API. Furthermore, the performance enhancement as achieved by both optimal PA and joint PA-TA over the fixed allocation increases with increasing API severity parameter $\Delta$ values. \textcolor{black}{To further investigate the relative impact of the amplitude and phase errors on the average stored energy degradation, in Fig.~\ref{fig:Amp_Phi} we have plotted the variation of the average stored energy at $\mathcal{U}$ due to hybrid EBF at $\mathcal{S}$ with  $N=20$ and $d=\{10,20\}$m for increasing degradations $\Delta_g$ and  $\Delta_\phi$ due to amplitude and phase errors, respectively. The numerical results in Fig.~\ref{fig:Amp_Phi} suggest that the amplitude errors lead to more significant performance degradation as compared to the phase errors. With the exact stored energy performance being noted from these three dimensional contour plots, we  also notice that the degradation in performance becomes more prominent for larger RFET ranges $d$.}

\begin{figure}[!t] 	
	\centering\includegraphics[width=3.45in]{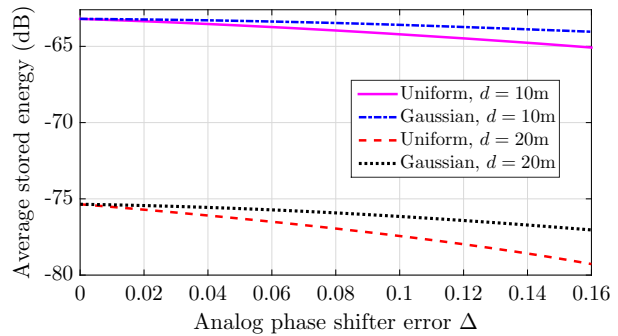} 
	\caption{\textcolor{black}{Comparing average stored energy  under  uniform and Gaussian distributions based modeling of API randomness.}}
	\label{fig:Gaus_Uni}   
\end{figure}

\textcolor{black}{Next, to investigate the impact of different random distributions for modeling API, we have considered a Gaussian distribution based API model here and numerically compared its impact on the average stored energy performance  in Fig.~\ref{fig:Gaus_Uni} against the uniform distribution model used in this work. Here, for fair comparison, keeping in mind the practical limitation on the values of API parameters, along with the fact that $>99.7\%$ of values in a Gaussian distribution lie within an interval of six standard deviations width around the mean, we have modeled API parameters as zero mean real Gaussian random variable with its standard deviation $\sigma_{\Psi}$ satisfying $3\sigma_{\Psi}=\frac{\Delta}{2}$. Hence, $\Psi\sim\mathbb{N}\left(0,\frac{\Delta^2}{36}\right).$ Intuitively, for both the distributions, the average stored energy performance degrades with increasing $\Delta$ due to  the higher severity of API and this degradation is more significant for larger values of RFET ranges $d$. Furthermore, for both $d=10$m and $d=20$m, the Gaussian distribution leads to lesser degradation in average stored energy as compared to the uniform modeling for API because the former has lower variance $\frac{\Delta^2}{36}$  than the latter having variance as  $\frac{1}{12}\left[\frac{\Delta}{2}-\left(\frac{\Delta}{2}\right)\right]^2=\frac{\Delta^2}{12}$. Hence, in comparison to the Gaussian one, with the consideration of uniform distribution based API modeling, we investigated the more severe degradation and with the proposed jointly optimized CE protocol we try to minimize the performance gap between the fully digital and  single RF chain based hybrid EBF designs.}

\begin{figure}[!t]		
	\centering\includegraphics[width=3.45in]{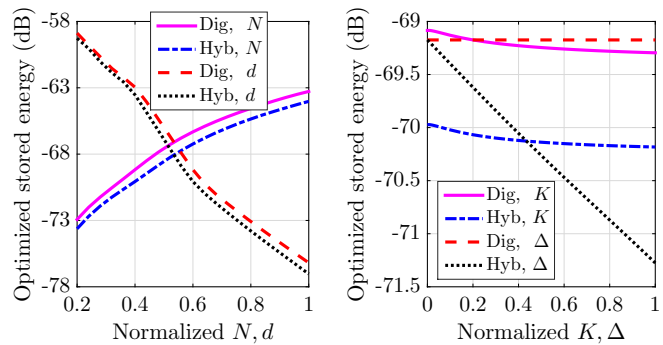} 
	\caption{Observing the trend in average stored energy at $\mathcal{U}$ due to digital EBF and the  proposed hybrid EBF protocols.}
	\label{fig:Dig-Hyb}  
\end{figure} 
Lastly, we compare the average stored energy performance for the jointly optimal PA-TA with proposed hybrid EBF having single RF chain against the conventional fully digital EBF protocol with $N$ RF chains~\cite{CSI-SU-WET,ICASSP18,Signal-WPT-TC17}. In Fig.~\ref{fig:Dig-Hyb}, this comparison is plotted for  different system parameters $\left(N,K,d,\Delta\right)$, whose values normalized to their respective maximum $\left(50,10,25\text{m},0.16\right)$ is shown. The joint PA-TA  for digital EBF can be obtained from \eqref{eq:jOPA} and \eqref{eq:jOTA}  defined for hybrid EBF, but with underlying $p_{c_i}$ and $\tau_{c_i}$ respectively replaced with $p_{c_i}^D$ and $\tau_{c_i}^D$, defined below.
\begin{eqnarray}\label{eq:opt-pci-D}
p_{c_i}^D\triangleq\frac{1}{\tau_c}\left[ \sqrt{\mathcal{A}_i\,p_d\left(\tau-\tau_c\right)\sigma_{\mathrm w}^{2}\left(N-1\right)} -\frac{\sigma_{\mathrm w}^{2}\left({K+1}\right)}{
	\beta}\right],
\end{eqnarray}  
\begin{eqnarray}\label{eq:opt-taui-D}
\tau_{c_i}^D\triangleq\frac{\sqrt{\frac{ \left(p_c \tau\,\beta+ {\sigma_{\mathrm w}^{2}\left({K+1}\right)}\right)\sigma_{\mathrm w}^{2}\left(N-1\right)}{\beta\left( \frac{\beta\,N}{K+1}+\norm{\boldsymbol{\mu}_{\mathbf{h}}}^2 +\frac{\mathcal{B}_i+p_c}{\mathcal{A}_i p_d}\right)}}-\frac{\sigma_{\mathrm w}^{2}\left({K+1}\right)}{
		\beta}}{p_c}.
\end{eqnarray}

We observe that with increasing $N$ and $d$, though  the performance gap between digital EBF and hybrid EBF increases, this gap is less than $1$dB. In contrast, this gap of about $1$dB remains almost invariant with increasing $K$. Moreover, this performance gap which is zero for $\Delta=0$ implying that under no API the proposed hybrid EBF can provide the exact performance as that of digital EBF under CE errors alone, increases with higher level of API severity  as incorporated by increasing $\Delta$ value. \textcolor{black}{Hence, we summarize that with general practical limits on the API parameter $\Delta<0.16$~\cite{PS1,PS2}, the average stored energy gap as achieved by  optimized  digital EBF and hybrid EBF under API and CE errors over Rician channels with nonlinear EH model at $\mathcal{U}$ is mostly less than $1$dB, which is very much acceptable practically given the high monetary cost and space constraints for having $N$ RF chains, especially for MISO systems with $N\gg1$.} 

\begin{figure}[!t] 	
	\centering\includegraphics[width=3.45in]{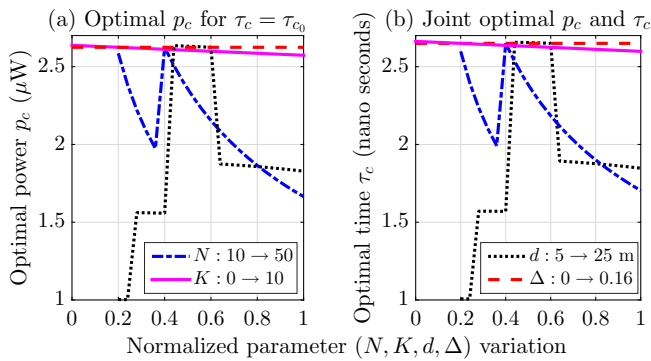}
	\caption{Insights on optimal PA and TA, with (a) plotting  optimal PA $p_c^*$ for $\tau_c=\tau_{c_0}$, and  (b) depicting optimal TA $\tau_{c_J}^*$ as returned by joint optimization algorithm with $p_c=p_{\max}$.}
	\label{fig:OPA-TA}  
\end{figure} 
\begin{figure}[!t] 	
	\centering\includegraphics[width=3.45in]{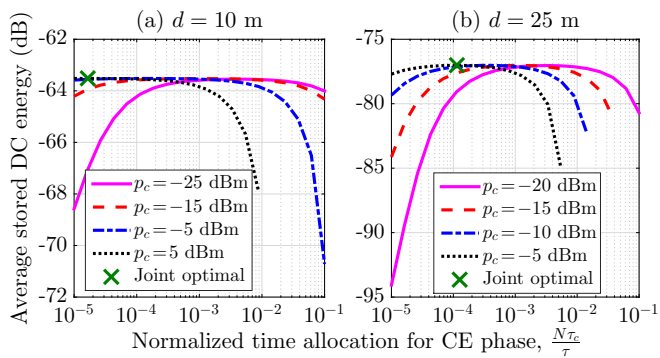}
	\caption{Insights on optimal TA $\tau_c$ with varying PA $p_c$ for $d=\{10,25\}$m. Maximum $\widehat\mu_{\mathrm{E}_s}^*$ as returned by the joint optimal design defined via Theorem~\ref{th:GOS} is also plotted.}
	\label{fig:OTA} 
\end{figure}  

\subsection{Insights on Optimal Power and Time Allocation}
This section focuses on bringing out the nontrivial design insights for the proposed hybrid EBF protocol. Specifically, in Fig.~\ref{fig:OPA-TA}(a), we depict the variation optimal PA $p_c^*$ for fixed TA $\tau_c=\tau_{c_0}$ (cf. Lemma~\ref{lem:OPA}) with different system parameters. Likewise, the corresponding trend or nature of the optimal TA $\tau_{c_J}^*$ with $ p_{c_J}^*=p_{\max}$ for the joint design, defined in Theorem~\ref{th:GOS}, is plotted in Fig.~\ref{fig:OPA-TA}(b). Interestingly, the optimal PA, solution of  $\mathcal{OP}1$, and optimal TA in joint design, solution of  $\mathcal{OP}$, follow a  similar tend for the variation of different parameters. The reason behind this being their similar effect on the energy consumption $\mathrm{E}_c=N\,p_c\tau_c$ at $\mathcal{U}$ during the CE phase, and the optimal PA being a constant $ p_{c_J}^*=p_{\max}$ in the joint design. Here, we notice that the optimal PA and  TA are independent of the variation in API parameter $\Delta$. However, their monotonically decreasing trend in $K$ is very slow. In contrast, the increase in the values of $N$ and $d$ has a stronger impact on the optimal solutions, which also gets affected by the nonlinear RF EH model. The sharp changes in the trend, as followed by optimal PA and TA, are due to the adopted PWLA model which has different linear definition between the two thresholds. Though, there is no clear trend in optimal PA and TA for varying $N$ and $d$, it can be observed that the optimal PA and TA in general decrease with increasing $N$. In contrast, these solutions first increase, and then decrease with the higher values for communication range  $d$. \textcolor{black}{Here, it may be recalled that the  CE time $\tau_c$ optimization for each sub-phase is very critical to overcome the limitations of having only one RF chain for estimating $N$ elements of the channel vector. As we have set $\tau=10$ms and $\tau-N\tau_c$ dedicated for RFET, we note that the optimal CE time for each sub-phase lies between $0.00001\%$ (for $N=50$) and $0.00003\%$ (for $N=10$) of the total coherence block duration $\tau$. Hence, we conclude that the proposed optimal antenna switching based CE is very efficient because the  TA $\tau_c$ for estimating the channel between $\mathcal{U}$ and each antenna element at $\mathcal{S}$ is negligible $\left(\approx10^{-7}\tau\right)$ in comparison to the coherence block duration $\tau$.}

To further gain key insights on the joint design, we plot the variation of the average harvested DC energy $\widehat\mu_{\mathrm{E}_s}$ with TA $\tau_c$ for different PA $p_c$ and $d$ in Fig.~\ref{fig:OTA}. It is observed that for larger $d$, implying weaker $\mathcal{S}$-to-$\mathcal{U}$ link quality, longer TA should be provided for the CE phase to obtain a better LSE of the underlying channel for  maximizing the achievable array gains. Further, with increasing TA for the CE phase, the optimal PA decreases to minimize the energy consumption needed in obtaining a desired LSE quality. Also, from the $\widehat\mu_{\mathrm{E}_s}^*$ achieved with jointly optimal PA and TA, as plotted in Fig.~\ref{fig:OTA} using `$\times$' marker we can verify the result in Theorem~\ref{th:GOS}, stating that setting PA to $p_c=p_{\max}$, while optimizing TA, results in the highest stored energy at $\mathcal{U}$.

\begin{figure}[!t]	 	
	\centering\includegraphics[width=3.45in]{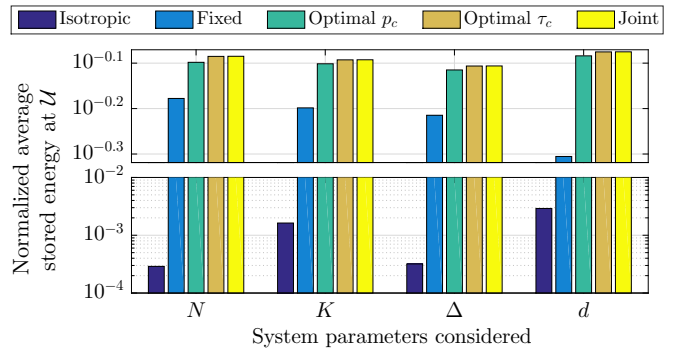} 
	\caption{Comparing the average stored energy, normalized to their respective maximum value  $\widehat\mu_{\mathrm{E}_s,\mathrm{id}}$, for the different precoder designs and system parameters.}
	\label{fig:comp1} 
\end{figure}  
 
\subsection{Performance Comparison and Achievable  EBF Gains} 
Now via Fig.~\ref{fig:comp1}, we conduct a performance comparison study among the five practical scenarios for varying  system parameters $N,\, K,\,d$, and $\Delta$ (cf. Figs.~\ref{fig:N} to~\ref{fig:API}). The performance comparison metric $\frac{\widehat\mu_{\mathrm{E}_s}}{\widehat\mu_{\mathrm{E}_s,\mathrm{id}}}$ considered here is the average stored energy $\widehat\mu_{\mathrm{E}_s}$ at $\mathcal{U}$ normalized to the maximum achievable energy $\widehat\mu_{\mathrm{E}_s,\mathrm{id}}$ under perfect CSI availability, given below:
\begin{align}
\mu_{E_h,\mathrm{id}}=&\;\mathbb{E}\left\lbrace \left(\tau-N\tau_c\right)\,p_d\,\left|\frac{\mathbf{h}^\mathrm{H}\,\mathbf{h}}{\norm{\mathbf{h}}}\right|^2-N\tau_c\,p_c\right\rbrace\nonumber\\
\approx&\; \widehat\mu_{\mathrm{E}_s,\mathrm{id}}\triangleq\left(\tau-N\tau_c\right)\,\mathcal{L}\left(\mu_{p_r,\mathrm{id}}\right)-N\tau_c\,p_c.
\end{align}
where \eqref{eq:ideal} is used along with the PWLA function  $\mathcal{L}\left(\cdot\right)$ as defined in \eqref{eq:PWLA0}.
Similarly using \eqref{eq:iso}, the average stored energy at $\mathcal{U}$ for the isotropic RFET  from $\mathcal{S}$ can be approximated as 
\begin{align}
\widehat\mu_{\mathrm{E}_s,\mathrm{iso}}\triangleq\left(\tau-N\tau_c\right)\,\mathcal{L}\left(\mu_{p_r,\mathrm{iso}}\right)-N\tau_c\,p_c.
\end{align} 
 
From Fig.~\ref{fig:comp1},  the average ratio $\frac{\widehat\mu_{\mathrm{E}_s,\mathrm{iso}}}{\widehat\mu_{\mathrm{E}_s,\mathrm{id}}}<0.0015$ implies that isotropic transmission is highly energy inefficient. On the other hand, the proposed joint optimal PA and TA scheme can help in achieving about $81.4\%$ of the maximum theoretically achievable performance, i.e., $0.814\,\widehat\mu_{\mathrm{E}_s,\mathrm{id}}$. Moreover, here the impact of optimal TA for fixed $p_c=p_{c_0}$ is much more significant, with an average performance of $0.8135\,\widehat\mu_{\mathrm{E}_s,\mathrm{id}}$, and it approximately reaches the performance achieved by jointly optimal PA and TA. The corresponding average performance of the optimal PA with fixed $\tau_c=\tau_{c_0}$ and fixed allocation $\left(p_c=p_{c_0},\tau_c=\tau_{c_0}\right)$ is approximately $0.8\,\widehat\mu_{\mathrm{E}_s,\mathrm{id}}$ and $0.6\,\widehat\mu_{\mathrm{E}_s,\mathrm{id}}$, respectively. This implies that optimal TA is a better semi-adaptive scheme and the joint PA-TA provides an average improvement of more than $37\%$ over the normalized stored energy performance of the fixed allocation scheme. 

\begin{figure}[!t]	
	\centering\includegraphics[width=3.45in]{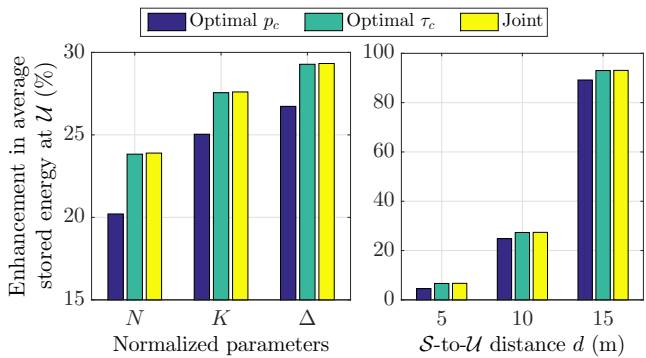} 
	\caption{Improvement in the average stored energy at $\mathcal{U}$ as achieved by different optimization schemes over the hybrid EBF with fixed allocations.}
	\label{fig:comp2}  
\end{figure}	 
Lastly, we quantify how this above mentioned average performance improvement achieved by the joint design over the fixed allocation varies for different system parameters values. From Fig.~\ref{fig:comp2} we observe that the optimal PA scheme  provides an average improvement of about $20\%$, $25\%$, and $27\%$ for the variation of $N$, $K$, and API parameter $\Delta,$ as respectively plotted earlier in Figs.~\ref{fig:N},~\ref{fig:K}, and~\ref{fig:API}. Whereas, it is slightly higher, viz., $24\%, 28\%,$ and $30\%$, respectively,  for the optimal TA scheme, which closely follows the performance of the jointly optimal design. Moreover, this enhancement gets much more  significant with increasing $\mathcal{S}$-to-$\mathcal{U}$ distance $d$ because the achievable analog EBF gains are strongly influenced by the wireless propagation losses. In fact, the EH performance improvement with joint design increases from about $6\%$ at $d=5$m,  to over $90\%$ at $d=15$m. Furthermore, at $d=20$m and maximum range $d=25$m, the jointly optimal design can respectively provide about $6$ times and $195$ times more stored energy $\widehat\mu_{\mathrm{E}_s}$ at $\mathcal{U}$ in comparison to that achieved with fixed PA-TA $\left(p_{c_0},\tau_{c_0}\right)$ scheme. This corroborates the utility of  proposed analysis and joint optimization  to enhance the practical  efficacy of the  hybrid EBF with single RF chain during  RFET over  Rician  channels under API and CE errors.

\section{Concluding Remarks}\label{sec:conclusion}  
\textcolor{black}{This work investigated the practical efficacy of using a single RF chain at a large antenna array power beacon in wirelessly delivering energy to a single antenna EH user $\mathcal{U}$ over Rician fading channels under practical API and CE errors.} Adopting a recently proposed EBF model that characterizes the real-world API, the optimal LSE for the effective channel is obtained. Next, using some practically-motivated tight analytical approximations for the key LSE-dependent statistics, the average energy stored   at $\mathcal{U}$ is derived in closed form while adopting a more refined nonlinear RF EH model. To maximize the achievable gains of the proposed hybrid EBF design, having lower cost and smaller form-factor, the optimal energy assignment at resource-constrained EH $\mathcal{U}$ is obtained via joint PA and TA for the CE phase global-optimally resolving the underlying CE-quality versus delivered-energy-quantity tradeoff. The proposed analysis has been validated by extensive simulations and these achievable gains over benchmark schemes have been numerically quantified. \textcolor{black}{Overall, the optimized hybrid EBF provides an average improvement of $37\%$ over fixed PA-TA, with a performance gap of less than $1$ dB as  compared to the digital EBF having $N$ RF chains. This corroborates the practical utility of the analysis and optimization carried out for the proposed hybrid EBF design incorporating API and nonlinear RF EH model. Hence, this investigation verifies that the smart hybrid EBF designs with single RF chain are indeed the practically promising solutions to closely realize  the maximum achievable array gains.} 

\textcolor{black}{In the future, we would like to extend the proposed hybrid CE protocol and optimized EBF design for serving multiple RF EH users by employing an optimal time-sharing policy among users to  solve the underlying CE-quality versus delivered-energy-quantity tradeoff.}  Another interesting direction includes the optimal pilot signal designing  for joint API compensation and CE with  longer training  duration to consider  dependence of API parameters over several coherence blocks. 

\bibliographystyle{IEEEtran} 

\end{document}